\documentclass[lettersize,journal]{IEEEtran}
\usepackage{amsmath,amssymb,amsfonts}
\usepackage{algorithmic}
\usepackage[ruled,vlined,linesnumbered]{algorithm2e}
\usepackage{array}
\usepackage[caption=false,font=normalsize,labelfont=sf,textfont=sf]{subfig}
\usepackage{pifont}
\usepackage{textcomp}
\usepackage{stfloats}
\usepackage{url}
\usepackage{verbatim}
\usepackage{graphicx}
\usepackage{multirow}
\usepackage{cite}
\begin{document}

\title{MOPAR: A Model Partitioning Framework for Deep Learning Inference Services on Serverless Platforms}

\author{Jiaang Duan, Shiyou Qian, ~\IEEEmembership{Member,~IEEE}, Dingyu Yang, Hanwen Hu, Jian Cao, ~\IEEEmembership{Senior Member,~IEEE}, Guangtao Xue, ~\IEEEmembership{Member,~IEEE}

\thanks{Manuscript received.}}

\markboth{IEEE TRANSACTIONS}%
{Shell \MakeLowercase{\textit{et al.}}: MOPAR: A Model Partitioning Framework for Deep Learning Inference Services on Serverless Platforms}

\IEEEpubid{0000--0000/00\$00.00~\copyright~2024 IEEE}

\maketitle

\begin{abstract}
With its elastic power and a pay-as-you-go cost model, the deployment of deep learning inference services (DLISs) on serverless platforms is emerging as a prevalent trend. However, the varying resource requirements of different layers in DL models hinder resource utilization and increase costs, when DLISs are deployed as a single function on serverless platforms. To tackle this problem, we propose a model partitioning framework called MOPAR. This work is based on the two resource usage patterns of DLISs: global differences and local similarity, due to the presence of resource dominant (RD) operators and layer stacking. Considering these patterns, MOPAR adopts a hybrid approach that initially divides the DL model vertically into multiple slices composed of similar layers to improve resource efficiency. Slices containing RD operators are further partitioned into multiple sub-slices, enabling parallel optimization to reduce inference latency. Moreover, MOPAR comprehensively employs data compression and share-memory techniques to offset the additional time introduced by communication between slices. We implement a prototype of MOPAR and evaluate its efficacy using four categories of 12 DL models on OpenFaaS and AWS Lambda. The experiment results show that MOPAR can improve the resource efficiency of DLISs by 27.62\% on average, while reducing latency by about 5.52\%. Furthermore, based on Lambda's pricing, the cost of running DLISs is reduced by about 2.58 $\times$ using MOPAR.
\end{abstract}


\IEEEPARstart{D}{eep} learning inference services (DLISs) are increasingly gaining traction as artificial intelligence demonstrates outstanding performance in various domains \cite{INFless}. Currently, cloud platforms host tens of thousands of DLIS instances to handle a substantial volume of user requests per second \cite{microsoft}. As the next-generation evolution of cloud platforms, serverless computing offers significant advantages such as elasticity, a pay-as-you-go cost model, and resource-free management \cite{BATCH, BARISTA}. 
Within this landscape, serverless inference, an extension of serverless computing, has emerged as a promising field that enables the cost-effective deployment of DL models for diverse applications, including Amazon Alexa\footnote{https://alexa.amazon.com/}, Facebook Messenger bots\footnote{https://developers.facebook.com/}.

Although the direct approach of deploying a DLIS in a function may seem efficient, it presents two drawbacks.
Firstly, resource allocation in this approach is based on the layers with the highest resource requirements, leading to reduced resource efficiency and increased costs \cite{shao2020communication}.
Secondly, the implementation of the direct approach becomes challenging as the size of DL models increases \cite{chatgpt}, 
since larger DLISs are more prone to generate resource fragmentation issues, which can significantly impact resource efficiency. Thus, the direct deployment approach is considered inefficient from the perspectives of both users and cloud providers.


Model partitioning (MP) is a promising approach to enhance resource utilization. It involves dividing a DL model into distributed instances, commonly referred to as model parallelism. MP has been extensively explored in both model training and inference \cite{wang2023topoopt, narayanan2021efficient}. 
Various \emph{throughput-oriented} MP methods have been proposed for model training, such as Baechi \cite{jeon2020baechi}, and Alpa \cite{zheng2022alpa}.
However, these approaches are not directly applicable to model inference due to the distinct computing mechanisms used in training. 
Furthermore, the impact of MP on model inference is gradually gaining attention. Tarnawski et al. \cite{tarnawski2020efficient} design an algorithm that splits the DNN into subgraphs per device to minimize latency. AlpaServe \cite{li2023alpaserve} is a system that incorporates model parallelism and placement to partition DLIS into several distinct slices to reduce latency. 
However, the applicability of AlpaServe to serverless environments is hindered by its enumeration strategy, resulting in increased latency and cost.

Meanwhile, some studies focus on enhancing the resource efficiency of DLISs deployed on serverless platforms. For example, BATCH \cite{BATCH} introduces a buffer layer that consolidates requests, thereby reducing costs through batching. Similarly, INFless \cite{INFless} leverages built-in batching and non-uniform scaling mechanisms to achieve low latency.
Nevertheless, these approaches still adopt the direct approach of deploying a DLIS in a function, disregarding the diverse resource requirements across distinct layers in a DLIS.

\IEEEpubidadjcol
By conducting a comprehensive evaluation of the trace data acquired from widely utilized DLISs and performing a feasible experiment, three significant insights are obtained  (\S \ref{sec:observation}).
Firstly, two crucial resource usage patterns of DLISs are identified, namely global differences and local similarities, due to the presence of resource dominant (RD) operators. RD operators lead to substantial variations in resource usage across different layers. 
Secondly, despite the introduction of communication costs, partitioning a DLIS into an appropriate number of slices can remarkably reduce the overall costs by conserving a substantial amount of computational resource.
Lastly, although the number of parameters in DLIS is typically substantial, their distribution remains constant. Hence, employing data compression techniques becomes feasible for minimizing the volume of data transmitted and reducing communication costs.

Therefore, the deployment of DLISs on serverless platforms in a fine-grained manner, aiming to enhance resource utilization and cost savings, is feasible. The resolution of this challenge entails tackling the following two inquiries. \textbf{Q1:} How can the MP strategy be efficiently employed to deploy DLISs incorporating diverse DL models on serverless platforms? Accomplishing an optimal partitioning of DLISs with minimal computational and communication expenses represents a complex challenge. \textbf{Q2:} How can the latency of DLISs be ensured while simultaneously reducing costs? Although there are advantages in utilizing MP, its employment also introduces additional latency that has the potential to offset those advantages.

To tackle these challenges, we propose a novel model partitioning framework called MOPAR (\S \ref{sec:design}).
Firstly, considering both global differences and local similarities, we employ a hybrid methodology that first vertically divides the DL model into multiple slices comprising analogous layers. This division enhances resource efficiency and cost savings. Subsequently, slices containing RD operators are further partitioned into multiple sub-slices, enabling parallel optimization to minimize latency.
Furthermore, we extensively leverage data compression techniques and share-memory mechanisms to mitigate the additional time incurred by communication between slices.

We implemented the prototype of MOPAR on two serverless platforms: the open-source serverless platform OpenFaaS\footnote{https://www.openfaas.com/} and the widely used platform Lambda (\S \ref{sec:eva}). Extensive experiments were conducted to evaluate the efficacy of MOPAR based on four categories of 12 DL models, including four convolutional neural networks (CNNs), two recurrent neural networks (RNNs), two graph neural networks (GCNs), and four Transformer-based models.
The experiment results show that MOPAR significantly enhances the resource efficiency of the eight non-transformer DLISs by 27.62\% on average, compared with five state-of-the-art baselines. Furthermore, by employing systematic data compression and communication optimization techniques, MOPAR effectively reduces the latency of DLISs by an average of 5.52\%. In the case of Lambda, MOPAR achieves a 2.58$\times$ reduction in operating costs compared to the unsplit method.

Our contributions can be summarized as follows:
\begin{itemize}
\setlength{\parskip}{-1pt}

\item We identify significant resource usage patterns of DLISs, laying the foundation for designing fine-grained solutions to deploy DLISs on serverless platforms. 

\item We propose a hybrid partitioning framework named MOPAR specifically designed for DLISs on serverless platforms, aiming to optimize resource efficiency while ensuring service latency.

\item We implement MOPAR on OpenFaaS\footnote{The code of MOPAR can be accessed anonymously at: https://anonymous.4open.science/r/} and Lambda, and evaluate its performance through intensive experiments.
\end{itemize}


\section{Background and Observations}
\label{sec:observation}
In this section, we elaborate the fundamentals of DLISs and analyze the inefficiency of deploying a DLIS in a function on serverless platforms. Furthermore, we underscore the necessity of model partitioning. 

\subsection{Background}
\label{sec:background}
\textbf{Deep Learning Inference Services.} The widespread adoption of Model-as-a-Service (MaaS) can primarily be attributed to the transformative influence of cloud computing on the generation, administration, and utilization of computer resources \cite{MaaS}. MaaS offers a diverse range of pre-trained models that can be easily customized for various applications. For example, Amazon SageMaker delivers a platform for the development, training, and deployment of DLISs\footnote{https://aws.amazon.com/sagemaker/}.

\begin{figure}[bt]
    \centering
    \includegraphics[width=1.0\linewidth]{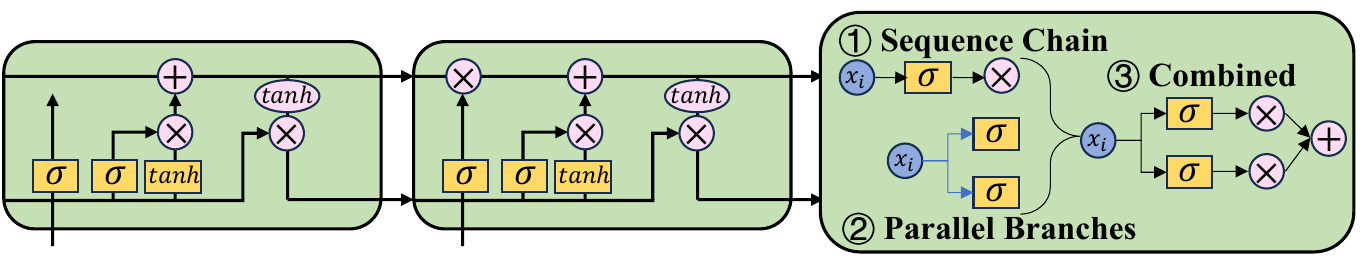}
    \caption{The structure of operators in LSTM}
    \label{fig:lstm_ope}
    \vspace{-5mm}
\end{figure}

\noindent
\textbf{Operators in DLIS.} The underlying model of a DLIS is organized into a series of consecutive layers, each comprising multiple operators. Within a DLIS, dominant operators are those that significantly contribute to the overall computational workload \cite{INFless}. Layers containing these dominant operations are referred to as RD layers. 
The interdependence among dominant operators within a layer can be formally represented by a directed acyclic graph (DAG) $G = <V, E>$, where $V$ represents the set of dominant operators and $E$ denotes the operator dependencies. As depicted in Fig. \ref{fig:lstm_ope}, the DAG of layers can be classified into three topologies: sequential chain, parallel branch, and a combination of both. 
Although these three basic topologies can constitute DLISs with diverse structures, their dominant operators are common. As discussed in \cite{INFless}, a comprehensive analysis of 11 popular models including VGG \cite{VGG16}, ResNet \cite{resnet}, LSTM \cite{LSTM-2365} and Bert \cite{bert} reveals the presence of only 71 different operators.


\noindent
\textbf{DLISs on Serverless Platforms.}
The serverless architecture is emerging as the next phase of cloud computing \cite{Serverless}.
Serverless computing is a promising solution for DLISs, such as SageMaker\footnote{https://aws.amazon.com/sagemaker/}. Currently, each DLIS is commonly encapsulated as a function on serverless platforms (e.g., a Docker container\footnote{https://www.docker.com/}). The number of function instances adjusts automatically according to service requests.

\begin{figure*}[tb]
  \centering
  \subfloat[EfficientNet]{
    \includegraphics[width = 0.23\linewidth]{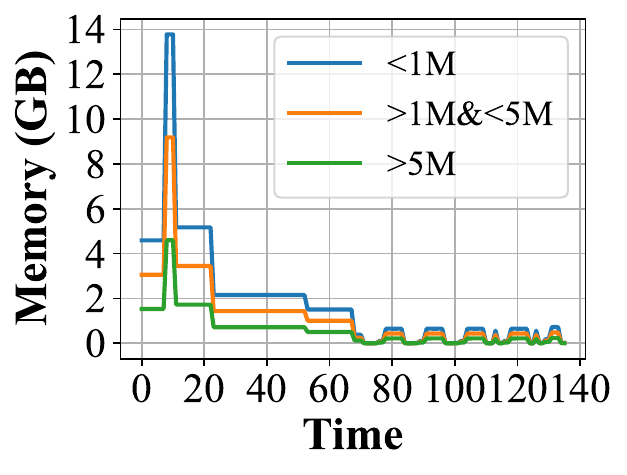}
        \label{fig:3/Mobilenet}
    \vspace{-3mm}
    }
  \subfloat[ConvNeXt]{
    \includegraphics[width = 0.244\linewidth]{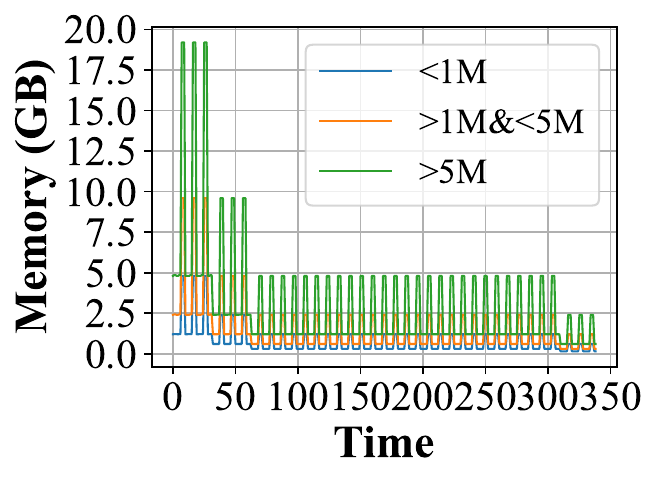}
        \label{fig:3/Transformer}
    \vspace{-3mm}
  }
  \subfloat[Cumulative memory spent]{
    \includegraphics[width = 0.23\linewidth]{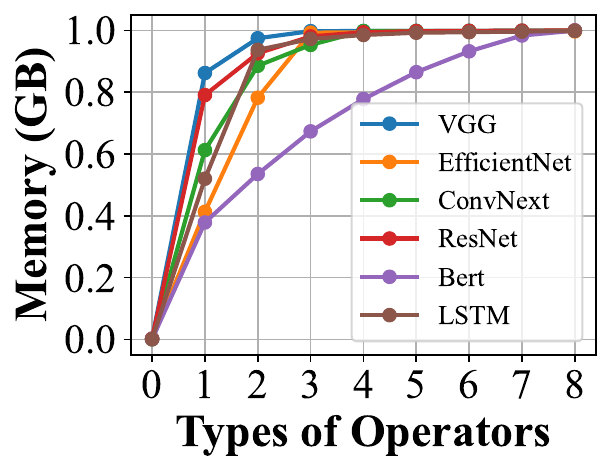}
        \label{fig:Cumulative Memory Spent}
    }
  \subfloat[Cost of different slices]{
    \includegraphics[width = 0.23\linewidth]{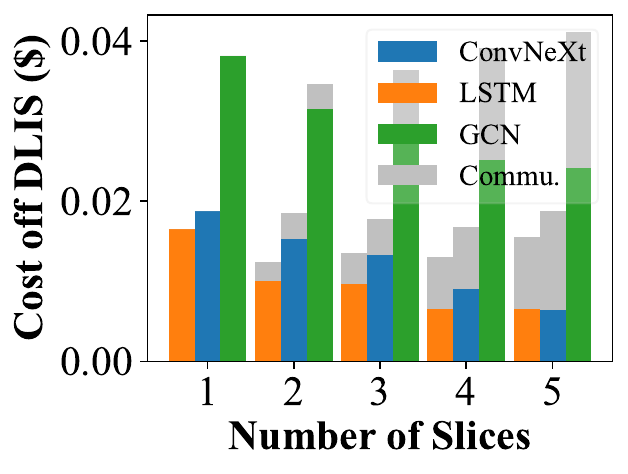}
    \label{fig:cost}
    }
    \vspace{-2mm}
\caption{Statistics regarding the consumption of resources, dominant operators, and costs associated with DLISs.}
\label{fig:Memory Footprint of Inference Model Submitted by Different Input}
\vspace{-4mm}
\end{figure*}

\subsection{Observations}
\label{sec:obser}
We conduct a study involving two distinct datasets. Firstly, we run experiments to profile the memory footprint of each layer in EfficientNet\footnote{https://pytorch.org/hub/nvidia\_deeplearningexamples\_efficientnet/}, and ConvNeXt \cite{convNeXt} models with varying input sizes.\footnote{Based on common resource allocation schemes in commercial serverless platforms such as AWS and Google, we encapsulate each DLIS as a function with a designated memory size in the experiments. The number of CPUs/vCPUs is determined proportionally according to the memory footprint. A function instance is scheduled to respond upon receiving a service request.} Section \ref{sec:eva} provides further details on the experimental setup. Secondly, we extract a wide variety of DLISs from the platform for artificial intelligence (PAI) over a period of 60 days. These DLISs encompass various domains, including advertising, shopping, and autoresponder bots \cite{weng2022mlaas}.
By analyzing the characteristics of DLISs, we identify certain limitations associated with the direct approach of encapsulating each DLIS as a function on serverless platforms.

\emph{\textbf{Observation $1$:} The resource consumption of DLISs deployed in individual functions exhibits significant fluctuations over their lifespan due to the presence of RD operators, thereby resulting in suboptimal resource utilization.}

The results of our analysis, as depicted in Fig. \ref{fig:3/Mobilenet} and \ref{fig:3/Transformer}, reveal two crucial resource usage patterns of DLISs: global differences and local similarities. 
Firstly, the memory footprint of DLISs fluctuates significantly. Specifically, during the execution of EfficientNet and ConvNeXt, the memory utilization experiences fluctuations of up to 37.52\% and 64.31\% respectively. These global differences stem from the different complexity of arithmetic operations performed across different layers within DL models. 
Secondly, DLISs exhibit the feature of resource usage similarity in neighboring layers. For instance, both ConvNeXt and EfficientNet display local similarities. This can be attributed to the common practice of stacking multiple layers with an equal number of parameters in DLISs, which enhances nonlinearity while simultaneously avoiding the accumulation of an excessive number of parameters within an individual layer.

Moreover, the resource usage of DLISs is determined by the RD operators employed in the underlying model.
By conducting an analysis of eight widely adopted DL models, as depicted in Fig. \ref{fig:Cumulative Memory Spent}, we make two findings: 
1) DL models typically exhibit a limited range of operator types, usually fewer than 30, due to the stacking of diverse layers in DL models, such as $Conv$ layers in VGG and self-attention in BERT.
2) Resource consumption is primarily governed by a small subset of RD operators. For example, LSTM-2365 \cite{INFless} contains 35 operators, with the $MatMul$ operator accounting for 76\% of the latency and 90\% of the memory footprint. Similarly, more than 95\% of the latency and resource consumption in ResNet-50 is attributed to the $Conv2D$ operator.

\emph{\textbf{Observation $2$}:
While model partitioning does result in extra communication expenses, it still stands as a feasible approach for effectively reducing overall costs.}

The overall execution cost of DLISs primarily revolves around the aggregate costs related to computation and communication. When a single DLIS is divided into multiple slices, the transmission of intermediate tensors is necessary between slices, thereby incurring complementary communication costs. With regard to the pricing pertaining to resource utilization and network communication in AWS Lambda\footnote{https://aws.amazon.com/lambda/}, we calculate the average computational and communication costs of four DLISs. In Fig. \ref{fig:cost}, we assess the costs associated with various model partitioning schemes. When the number of partitioned slices is small, the computational cost accounts for a larger proportion of the total cost since DLISs are typically both computation and memory intensive. Despite the introduction of communication costs resulting from model partitioning, a significant amount of computational cost can be conserved. For example, upon dividing the ConvNeXt model into three slices, the communication cost increases by 18.61\%, while the computing cost is reduced by 36.43\%. 
However, with an increasing number of slices, the communication cost becomes increasingly prominent, gradually diminishing the advantages brought about by model partitioning.

Furthermore, it can be observed that the escalation in communication costs is not directly proportional to the number of slices. For instance, whenever an additional slice is incorporated into ConvNeXt, the communication costs increase disproportionately. In addition, compared with ConvNeXt with four slices, the communication costs escalate by 61\% when five slices are included, while increasing slices from 2 to three leads to a 30.6\% increase. This discrepancy arises from the fact that communication costs depend on the volume of data transferred between slices. The greater the volume of data transmission, the higher the communication cost. Hence, it is imperative to judiciously select the appropriate number of slices and split points in order to reduce the cost of DLISs.


\begin{figure}[tb]
    \centering
    \vspace{-3mm}
     \includegraphics[scale=0.66]{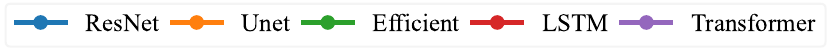}
     
    \vspace{-5mm}  
    \subfloat[Communication spent ]{
    \includegraphics[width = 0.48\linewidth]{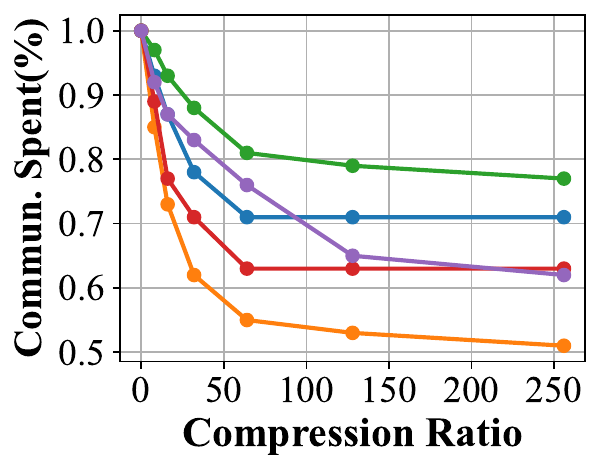}
    \label{fig:compression}
  }
    \subfloat[Performance loss]{
    \includegraphics[width = 0.5\linewidth]{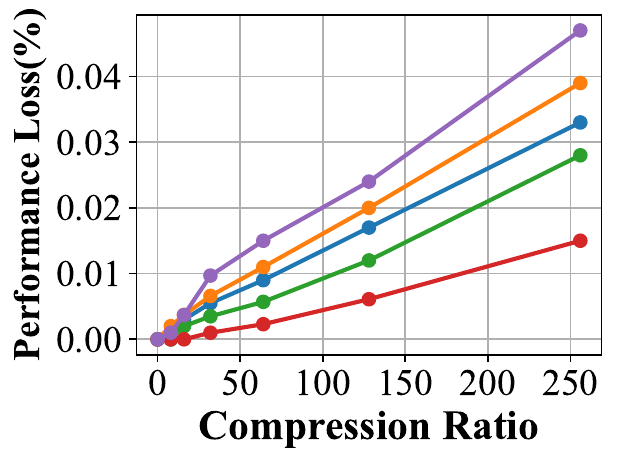}
    \label{fig:performance}
    }\vspace{-2mm}
\caption{Costs and performance with different compression ratios}
\vspace{-4mm}
\label{fig:cost factors}
\end{figure}
\emph{\textbf{Observation $3$}:
Data compression can significantly enhance the efficiency of communication while having minimal impact on the performance of DLISs.}

We investigate the impact of data compression on the performance and cost of five DLISs, including ResNet, Unet, Efficient, LSTM, and Transformer. We partition each model into two slices and compare the communication costs and performance with/without data compression. We can gain a comprehensive understanding of the influence of data compression on DLISs, as shown in Fig. \ref{fig:cost factors}.

Firstly, the ratio of data compression directly influences cost reduction. For instance, Fig. \ref{fig:compression} demonstrates that Transformer can achieve an 8\% reduction in communication costs with a compression ratio of 8, and a 24\% reduction with a ratio of 64. As data is compressed, the transferred data volume can be minimized, leading to significant cost savings. Therefore, the communication cost incurred by MP in Fig. \ref{fig:Memory Footprint of Inference Model Submitted by Different Input} can be further diminished through data compression.

Secondly, there is a threshold beyond which further data compression does not result in considerable cost reduction. For instance, Fig. \ref{fig:compression} clearly illustrates that when the compression ratio exceeds 64, the costs of ResNet and LSTM no longer decrease. This phenomenon can be attributed to the Shannon-Hartley theorem \cite{rioul2014shannon}, which states that there is a maximum limit to the amount of information that can be transmitted over a limited bandwidth channel. Once the compression ratio reaches this limit, any additional compression will not be effective in reducing the amount of data transmitted.

Thirdly, Fig. \ref{fig:performance} provides insights that as the compression ratio increases, there is a decline in the performance of DLISs. For instance, when the compression ratio is set to 256, Unet exhibits a performance loss of 0.04\%. This can be attributed to the fact that higher compression ratios often lead to an increase in information loss. 
Therefore, it is crucial to carefully consider the trade-off between compression ratio and model accuracy.

\subsection{Motivation}

Although serverless computing offers a granular resource management paradigm, the direct encapsulation of a DLIS as a function leads to low resource utilization and increased costs due to the variations in resource requirements across different layers within DLISs. 
Fortunately, MP can effectively reduce costs by dividing a DLIS into multiple slices, with each slice deployed as an individual function. MP offers optimization opportunities by enabling resource allocation refinement. However, the communication overhead between slices introduces additional latency, posing a challenge to achieve cost savings for DLISs while ensuring low latency.


\section{The Design of MOPAR}
\label{sec:design}
\subsection{Overview}

MOPAR is a model partitioning framework that divides a DLIS into multiple slices, each encapsulated as a serverless function with lower resource configurations. When designing MOPAR, we consider two objectives. On the one hand, model partitioning based on DLIS resource usage patterns can enhance resource efficiency and lead to cost savings. On the other hand, serving as online services, DLISs have strict service-level objectives (SLO) in terms of latency. Due to the introduction of additional communication between slices, model partitioning inevitably increases the latency of DLISs. Therefore, MOPAR endeavors to strike a tradeoff between these two aspects.

\begin{figure}[tb] 
\begin{center}
\includegraphics[width=\linewidth,scale=1.0]{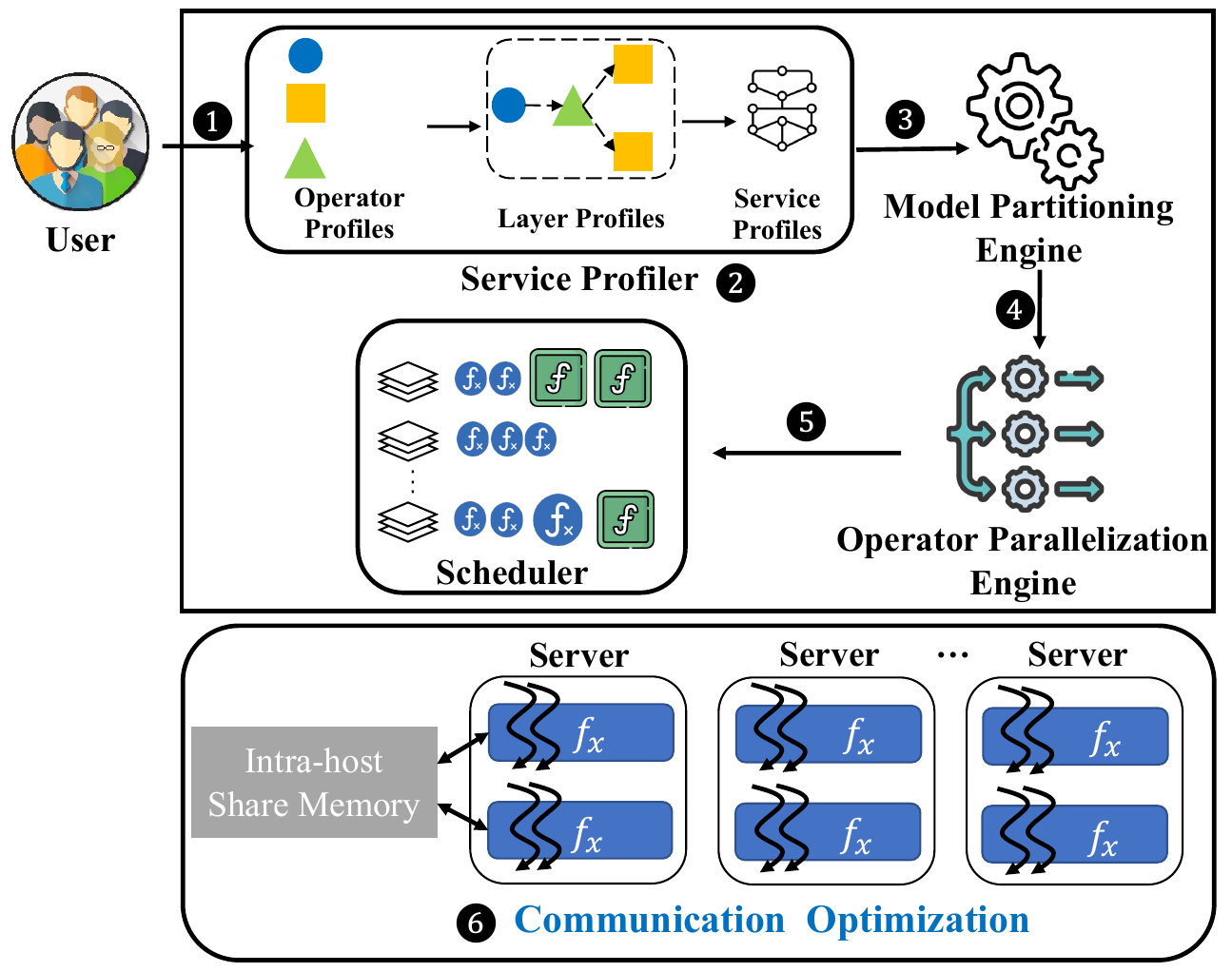}
    \caption{The architecture of MOPAR}
    \label{fig:FaaS_structure}
\end{center}
\vspace{-6mm}
\end{figure}

The architecture of MOPAR is shown in Fig. \ref{fig:FaaS_structure}, which consists of three components: a service profiler (SP), a model partitioning engine (MPE), and a communication optimization module (COM). The MPE serves as the central element of MOPAR, generating a hybrid model partitioning scheme based on model profiles and parallelizing slices containing RD operators to reduce their execution time. Furthermore, the COM extensively employs compression techniques and share-memory communication to minimize latency. The combination of the MPE and COM effectively compensates for the additional latency introduced between slices.

The workflow of MOPAR processing requests is also presented in Fig. \ref{fig:FaaS_structure}, which consists of two primary stages. During the preparation stage, for each user-submitted DLIS request (\ding{172}), the resource footprint and latency of the corresponding model are estimated by the SP (\ding{173}). Based on the model profile, the MPE generates a partitioning scheme (\ding{174}), and configures resource allocation and parallelization settings (\ding{175}). In the deployment stage, the slices derived from a model are scheduled in the cluster to minimize overall costs (\ding{176}). After scheduling, the share-memory technique is employed to facilitate data exchange between slices (\ding{177}).

\subsection{Service Profiler}
\label{sec:sp}

As explicated in Section \ref{sec:background}, a DLIS consists of a series of consecutive layers, each comprising various operators that execute specific computations on the input data. Accordingly, MOPAR adopts a three-level profiling method, which focuses on operators, layers, and overall service. At the operator level, MOPAR performs profiling based on input size to capture the memory footprint and execution duration of the dominant operators, as they typically consume the majority of resources and time. At the layer level, MOPAR derives the memory footprint and execution time of the layer by consolidating the profiles of its constituent operators. 
At the service level, MOPAR leverages the layer profiles to construct a concise vector representation of the overall service profile.

\noindent
\textbf{Operator Profile.} 
\label{sec:op_pro}
As stated in Section \ref{sec:background}, the profiling of DLISs primarily revolves around the identification and analysis of dominant operators. Typically, a DLIS contains less than 15 dominant operators as discussed in \cite{INFless}.
Analysis reveals that the impact of dominant operators on the resource utilization of DLIS is significant, accounting for over 80\%. 
Hence, MOPAR arranges the memory footprint of operators in each layer in ascending order and chooses the top $n$ operators responsible for 80\% of the memory consumption in that layer as the dominant operators. DLISs usually undergo repeated invocations, generating a substantial volume of trace data. Therefore, MOPAR adopts a data-driven profiling strategy. 


Given a dominant operator $O_i$, let $X$ be the model to which the operator belongs, $s$ be the input size (e.g., the size of an image), $p$ denote the number of layer parameters, and $m_i$ and $t_i$ represent the memory footprint and execution duration of $O_i$, respectively.
MOPAR establishes a prediction model to characterize the relationship between $<X, s, p>$ and $<m_i, t_i>$ for each dominant operator. 
Note that $X$, $s$ and $p$ are three independent variables, whereas $m_i$ and $t_i$ are dependent variables. Inspired by \cite{Horus}, we use linear regression (LR)\footnote{https://en.wikipedia.org/wiki/Linear\_regression}, XGBoost\footnote{https://xgboost.readthedocs.io/en/stable/}, and random forest (RF)\footnote{https://en.wikipedia.org/wiki/Random\_forest} to establish the prediction model for the dominant operators.

\noindent
\textbf{Layer Profile.}
Upon acquiring the memory footprint and execution time of the dominant operators within a layer, the layer profile can be aggregated according to the layer topology.

i) For sequential chain topology, the layer memory footprint $M_c$ is the maximum of the memory footprint of all the constituent operators in the chain, and the layer execution time $t_c$ is the sum of all the operators, defined as:
\begin{equation}
\vspace{-4mm}
  \left\{
    \begin{aligned}
    & M_{c} = \max_{o_i \in Layer} m_i \\
    & t_{c} = \sum_{o_i \in Layer} t_i  
    \end{aligned}
  \right.
\end{equation}

ii) For parallel topology with $n$ branches $B\!=\!\{b_1,b_2,\cdots,b_n\}$, the $j$-th operator on the $i$-th branch is denoted by $O_{ij}$ whose memory footprint and execution time are $m_{ij}$ and $t_{ij}$ respectively. In addition, the longest length among $n$ branches is denoted as $\kappa$. Since each operator at the same position on each branch can execute in parallel, the total memory footprint at the $j$-th position is $M_{b_j} = \sum_{i=1}^n m_{ij}$. Therefore, the layer memory footprint $M_b$ is the maximum at all positions.
The execution time at the $j$-th position is $t_{p_j}=\max(t_{1j},...,t_{nj})$. The layer execution time is the sum at all positions.
\begin{equation}
  \left\{
    \begin{aligned}
    & M_{b} = \max(M_{p_1},M_{p_2},\cdots,M_{p_\kappa}) \\
    & t_{b} = \sum_{j=1}^\kappa t_{p_j}  
    \end{aligned}
  \right.
\end{equation}


iii) For a hybrid topology combining sequential chains and parallel branches, we can estimate the overall memory footprint by taking into account both the requirements for the sequential chain part ($M_c$) and the parallel branches part ($M_b$). Specifically, the total memory footprint and execution time can be approximated as:
\begin{equation}
  \left\{
    \begin{aligned}
    & M_{h} = \max\{M_{c},M_{b}\} \\
    & t_{h} = t_{c}+t_{b}
    \end{aligned}
  \right.
\end{equation}

\noindent
\textbf{Service Profile.} For a model with $L$ layers, based on the memory footprint of each layer, the profile of DLIS running the model is represented as:
\begin{equation}
  \left\{
    \begin{aligned}
    &  \textbf{M} = \{M_1, M_2, ..., M_L\} \\
    &  \textbf{T} = \{t_1,t_2,\cdots,t_L\}
    \end{aligned}
  \right.
\end{equation}


\noindent
\textbf{Prediction Evaluation}. Accurately predicting the memory footprint of DLISs is a critical requirement to ensure their efficient execution. While overestimation of the memory footprint may not be optimal for resource utilization, underestimation can result in service failures due to out-of-memory (OOM). 
Therefore, we train memory footprint prediction models based on the root mean square log error (RMSLE) metric. This metric imposes a heavier penalty on underestimation than overestimation, which is imperative to prioritize the mitigation of service failures caused by underestimating resource consumption.

\begin{figure}[htb]
  \vspace{-4mm}
  \centering
  \includegraphics[scale=0.4]{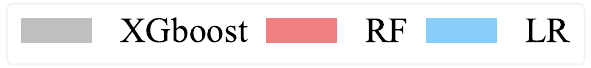}
    \vspace{-5mm}  
  
  \subfloat[Operator memory footprint]{
    \includegraphics[width = 0.48\linewidth]{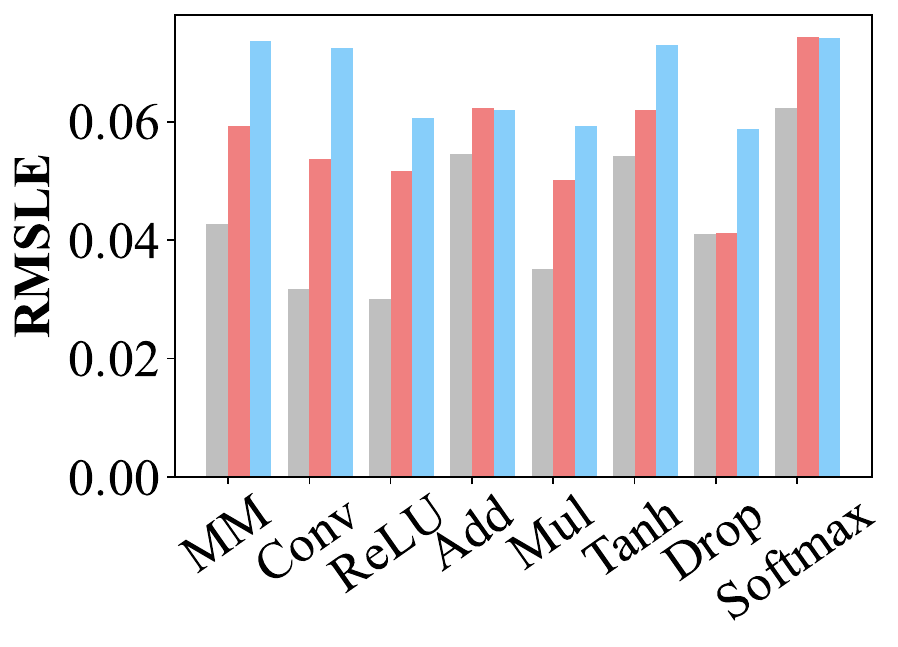}
        \label{fig:RMSLE score}
  }
  \subfloat[Operator execution time]{
    \includegraphics[width = 0.47\linewidth]{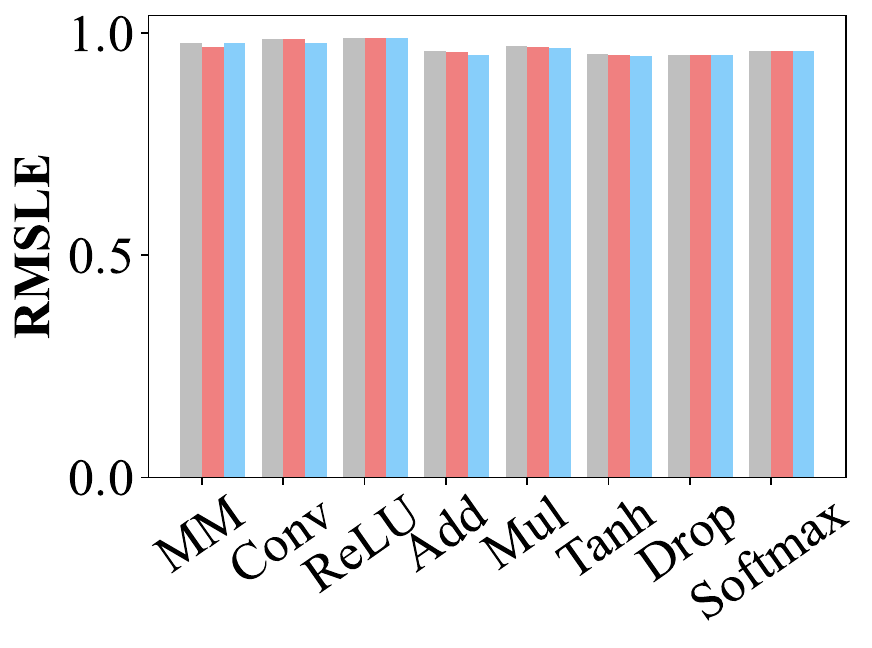}
    \label{fig:time accuracy}
  } 
\caption{The evaluation of three prediction methods for estimating the memory footprint and the execution time of operators}
\vspace{-4mm}
\end{figure}

We evaluate the performance in estimating the memory footprint of 118 operators across eight different models. These operators are collectively called 1600 times. Specifically, we select 1200 samples as the training set and use the remaining samples as the validation set. The results show that XGBoost outperforms the other two methods. The prediction results of eight representative operators are shown in Figure \ref{fig:RMSLE score}, with an RMSLE score of 0.0319 for $Conv2D$ and 0.0427 for $MatMul$. Similarly, regarding profiling layers, XGBoost achieves the lowest RMSLE score of 0.105, as illustrated in Table \ref{table:RMLSE}. 
Additionally, concerning the execution time, as dipiected in Fig. \ref{fig:time accuracy}, the average prediction accuracy of the three models is higher than 95\%. Therefore, we use XGBoost as the prediction model in MOPAR.


\vspace{3mm}
\begin{table}[tb]
  \centering
  \caption{RMSLE of layer memory footprint prediction}
  \begin{tabular}{c c c c} 
    \hline
       &\textbf{LR} & \textbf{XGBoost} &\textbf{RF}\\ [0.5ex]
    \hline
    RMSLE & 0.156 & 0.105 & 0.139  \\\hline
  \end{tabular}
  \label{table:RMLSE}
\end{table}


\subsection{Model Partitioning Engine}
\label{sec:hypad}
Observation 2 highlights the effectiveness of MP in cost savings for DLISs. 
As a result, the MPE in MOPAR is employed to partition the DLIS both vertically and horizontally. 
After vertical partitioning, MPE deploys each slice as a function with lower resource requirements. 
Subsequently, we are inspired by the classical model parallelism strategy of sharding a slice horizontally into multiple sub-slices, each of which contains a fraction of model parameters \cite{clockwork}. 

The primary objective of the MPE should be directed towards the enhancement of efficiency in resource utilization and cost savings, with the constraint on latency. Although the partitioning of DLIS can be advantageous in terms of improving the efficiency of resource usage, it is worth noting that an excessive number of slices may result in a significant increase in communication time between slices. Moreover, the degree of parallelism of slices can also influence latency. Consequently, it becomes imperative for the MPE to determine the optimal split points and the corresponding configurations of parallelism, while striking a suitable balance between resource usage efficiency and latency.

\noindent
\textbf{Optimization Objective.} 
The problem of partitioning a DLIS can be formalized as follows. 
First of all, based on the service profile, the DLIS can be expressed by a graph $G=<V, E>$, where each node $l_i \in V$ corresponds to a distinct layer within DLIS, and each edge $<l_i, l_j>\in E$ corresponds to a tensor serving as the output of layer $l_i$ and the input of layer $l_j$.
Moreover, a parallelization configuration $\gamma_i$ for a slice $i$ is applied to each layer in the slice when segmenting horizontally. 
Let $c_m$ be the cost per unit of the allocated memory for each slice, and $c_n$ be the unit price of network transmission.
Ultimately, we expect to obtain a hybrid partitioning strategy containing a vertical partitioning strategy $\pi_S$ and a horizontal parallelism strategy $\pi_P$ to optimize resource utilization while speeding up the execution time. 

To quantify the effect of $\pi_S$ and $\pi_P$, we define three indicators.
Firstly, $t_p (l_i,\gamma_i)$ is defined as the execution time of layer $l_i$ with parallelization configuration $\gamma_i$. 
Secondly, $t_a (l_i, \gamma_i)$ is defined to measure the time required for the parameter aggregation in layer $l_i$ with $\gamma_i$.
Thirdly, for a tensor $e$ transferring between layer $l_i$ with $\gamma_i$ and $l_j$ with $\gamma_j$, $t_c(e, \gamma_i, \gamma_j)$ is introduced as the communication time between $l_i$ and $l_j$, which can be computed based on data size and communication bandwidth. Hence, the total execution time $t_\tau$ of DLIS $\tau$ under a hybrid partitioning strategy ($\pi_S,\pi_P$) can be expressed as:
\begin{equation}
    \begin{aligned}
        t_\tau(G,\pi_S,\pi_P) &= \sum_{l_i \in V}(t_p(l_i,\gamma_i)+t_a(l_i,\gamma_i)) \\
        & + \sum_{<l_i,l_j> \in E} t_c(e,\gamma_i,\gamma_j)
    \end{aligned}
    \label{eq:time}
\end{equation}


Given a DLIS with $L$ layers, the vertical partitioning strategy $\pi_S$ is to search for $k$ split points $\{S_1, S_2, \cdots, S_{k}\}$ among these $L$ layers, where $1 \leq S_i \leq L$ and $S_i<S_j$ when $i<j$. These $k$ points partition the $L$ layers into $k+1$ slices $\Omega=\{g_1,g_2,\cdots,g_{k+1}\}$, where the memory allocated for each slice is denoted as $M_{g_i}$. Additionally, the parallelization strategy $\pi_P$ of each slice corresponding to $\pi_S$ is to cut each slice horizontally into $\eta$ equal-size sub-slices $\{P_1, P_2,\cdots, P_{\eta}\}$, satisfying $1 \leq \eta \leq \frac{M_{g_i}}{\lambda}$, where $\lambda$ is the ratio of memory to vCPUs usually specified by serverless platforms. 
Therefore, the objective of hybrid partitioning is to find the optimal partitioning strategy that minimizes costs for DLIS:

\begin{equation}
    \begin{aligned}
        &\min_{\pi_S} \sum_{g_i \in \Omega} (M_{g_i}*(t_p+t_a) * c_m) + 
        c_n * \sum_{<l_i,l_j>\in E}t_c\\
        &s.t. \left\{
            \begin{aligned}
                & \eta \leq \frac{M_{g_i}}{\lambda} \\
                & t_\tau(G,\pi_S,\pi_P) \leq \sum_{l_i \in G} (t_{l_i}) \\
                & c_m \geq 0, c_n \geq 0
            \end{aligned}
        \right.
    \end{aligned}
\end{equation}
where $t_{l_i}$ is the execution time of each layer before partitioning, which can be obtained from its service profile. The objective function takes into account the impact of parallelization on reducing the execution time. This allows for the utilization of MP optimize resource efficiency while ensuring that the total latency with MP does not exceed the original latency without MP, as defined by the second constraint.

\noindent
\textbf{Node and Edge Elimination.} 
To achieve the goal of cost optimization, it is crucial that the resource allocation for each layer within a slice is as uniform as possible.  
While layer-to-layer partitioning can maximize resource efficiency for each layer, it also incurs substantial latency due to the introduction of communication between slices. 
Furthermore, the number of potential strategies for parallelization increases exponentially as the number of layers in DLIS grows, making it unfeasible to develop a unique strategy for each layer.

Fortunately, based on Observation 1, the memory footprint of successive layers exhibits strong similarity, which gives us the inspiration to design our algorithm.
We use an elimination strategy to first combine multiple successive layers with similar resource footprints as a single slice and then configure different degrees of parallelism for each slice. This method ensures that the memory footprint of each slice is consistent while reducing latency. Note that the terms ``slice'' and ``group'', ``layer'' and ``node'' are interchangeable in this work.


\begin{figure}[tb]
  \vspace{-4mm}
  \subfloat[Node elimination]{
    \includegraphics[width = 0.52\linewidth]{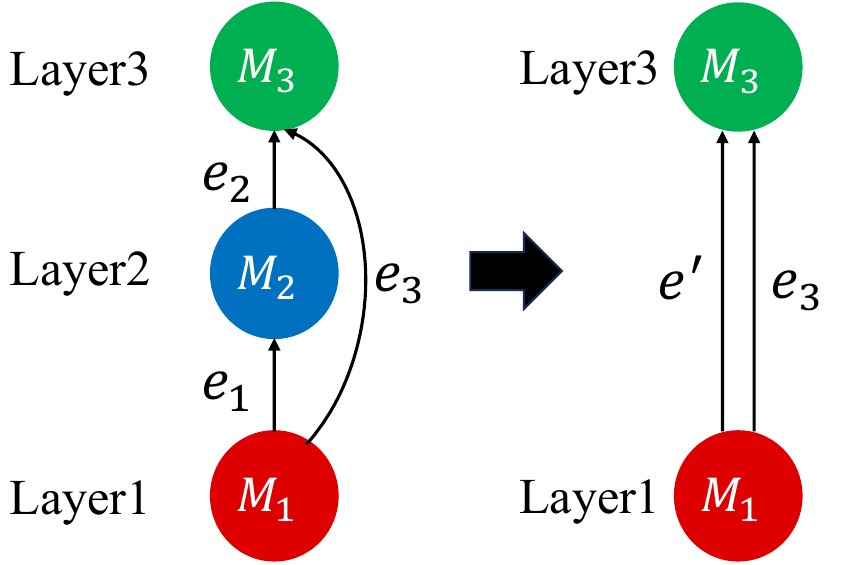}
        \label{fig:node_eli}
  }
  \subfloat[Edge elimination]{
    \includegraphics[width = 0.44\linewidth]{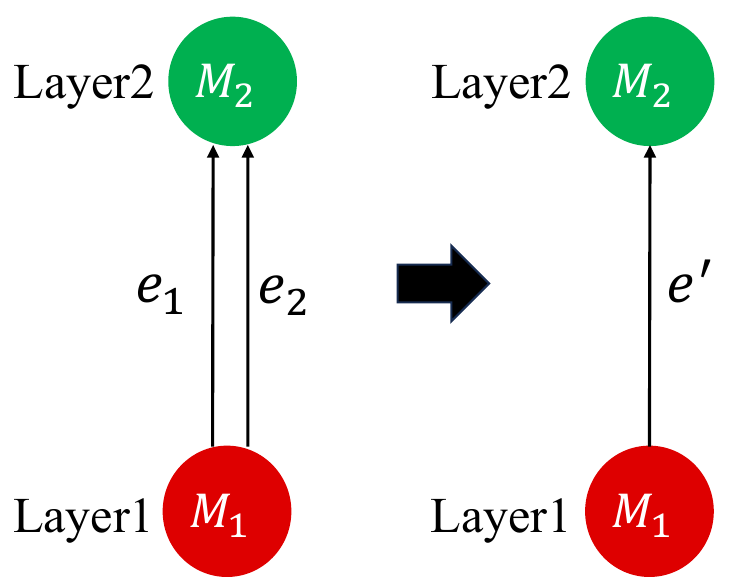}
    \label{fig:edge_eli}
  } 
\caption{Node and edge elimination for DLIS}
\vspace{-3mm}
\end{figure}

The specific elimination step involves two primary aspects. First, for layer $l_j$ in $G$, if it has a single input edge $e_1$ and a single output edge $e_2$, and the disparity in memory footprint between adjacent layers is within $5\%$ \footnote{The setting of the threshold primarily relies on the structure of underlying model and the accuracy of the prediction model.}, i.e., $\frac{|M_j-M_{j+1}|}{M_j}\leq 0.05$, the node elimination process removes layer $l_j$ and the two corresponding edges $e_1$ and $e_2$ are merged to introduce a new edge $e^{'}$, as shown in Figure \ref{fig:node_eli}. 
Second, due to node elimination, multiple edges may exist between two nodes, as depicted in Fig. \ref{fig:node_eli}. For two edges $e_1$ and $e_2$ with identical source and destination nodes, the process of edge elimination entails the removal of $e_1$ and $e_2$, followed by the insertion of a new edge $e^{'}$, as illustrated in Fig. \ref{fig:edge_eli}. 


\noindent
\textbf{Hybrid Partition Strategy.} 
Obtaining a model partitioning strategy primarily faces the following challenges: (1) Combination explosion: DL models are typically composed of multiple layers, each offering various partitioning options. Consequently, this gives rise to the issue of exponential growth in the number of partitioning solutions. (2) Local dependency: Dependencies may exist between different layers within the model, whereby the partitioning option of a specific layer may affect other layers.

By conducting a quantitative evaluation on the memory footprint and latency associated with different partitioning strategies for DLISs, we address the above challenges from two aspects. 
First of all, MPE simplifies the graph through node and edge eliminations to mitigate the issue of combination explosion.
Furthermore, MPE devises a dynamic programming (DP) algorithm to determine the vertical split points in the simplified graph. By considering local dependencies and calculating the optimal solution in stages, DP enables the determination of the global optimum. 
MPE then thoroughly explores all potential parallelism strategies for the simplified graph and sequentially determines the parallelism of each node to minimize its execution time. Lastly, MPE iteratively determines the parallelism of each layer in the original graph.

\begin{algorithm}[tb]
	\caption{Hybrid Partitioning Algorithm of DLISs}
	\label{alg:algorithm1}
        \KwIn{The DAG of a DLIS $G$, and profiling time $t_p$, $t_a$ and $t_c$, and ratio of memory to vCPUs $\lambda$}
    
	\KwOut{Set of split points $\textbf{S}$ and parallelism schemes of slices $\textbf{P}$}
   
    \textbf{Initialize} $G^{(0)} = G$, $\textbf{P} \gets \emptyset$, $\textbf{S} \gets \emptyset$; 
    
    
    \While{True}{
    $G^{(i+1)} \gets $ NodeElimination($G^{(i)}$)

    $G^{(i+2)} \gets $ EdgeElimination($G^{(i+1)}$)

    \If{ $G^{(i+2)} = G^{(i)}$}{
        $G^{'} = G^{(i)}$,
        \textbf{break}
    }
    }

    Count the number of nodes $n$ in $G^{'}$
    
        
        
    \For{$i \gets 1$ to $n$}{
        \For{$j \gets i$ to $n$}{
        
            $cost^{split}\gets cost^{com}_{[j]}+cost^{cal}_{[j - i]}$
            
            $dp_{[j]} \gets \min(dp_{[j]}, dp_{[i-1]} + cost^{split})$
        }
    }
    $i = n$
    
    \While{$i > 0$}{
        \For{$j \leftarrow i$ to $0$}{
            \If{$dp_{[i]} = dp_{[j-1]} + cost^{com}_{[i - 1]} + cost^{cal}_{[i - j]}$}{
            $\textbf{S} \gets \textbf{S}\cup i$

            $i \gets j - 1$
            
            \textbf{break}
            }
        }
    }
    \For{each node $v$ in $\textbf{S}$}{
        \For{$i \gets 1$ to $\frac{M_{v}}{\lambda}$ }{
        
            $\pi_P = \arg\min t_\tau(G,\pi_S,\pi_P)$
            
            $\textbf{P} \gets \textbf{P} \cup \pi_P$
        }    
    }

            

        
        
        
        
        
            

    
    
        
        
        
        
    

    \Return $\textbf{S}$ and $\textbf{P}$
\end{algorithm}

The hybrid partitioning algorithm for DLISs (HyPAD) is presented in Algorithm \ref{alg:algorithm1}, which consists of three steps: graph simplification, vertical partitioning strategy generation, and optimal parallelism strategy search. 
At step 1, HyPAD conducts iterative elimination of nodes and edges on the input graph $G$ of DLIS, resulting in a simplified graph $G'$ (lines 2-6). 
At step 2, HyPAD traverses each node in $G'$, calculating the cumulative computing cost $cost^{cal}_{i}$ and communication cost $cost^{com}_{i}$ for each layer $i$. Then, it employs DP to iteratively compute the minimum total cost, and subsequently performs backtracking to determine the optimal split strategy $\pi_S$ (lines 7-18).
Specifically, the $dp$ array is utilized to store the minimum total cost needed to reach each layer, with $dp[i]$ representing the minimum total cost from layer 0 to layer $i$, encompassing both communication and computing overhead. The values are updated during DP by comparing different split points, determining the optimal strategy for each layer. Ultimately, $dp[n]$ (where $n$ is the number of layers in the model) holds the minimum total cost for the entire model. In the backtracking phase, the split point array records the optimal split points for each layer, facilitating the determination of subsequent optimal configuration.

At step 3, HyPAD models the policy $\pi_{P}$ that minimizes the total time defined in Eq. (\ref{eq:time}) and explores all potential parallelism strategies for each layer in the simplified graph $G'$ (lines 19-23). 
Ultimately, the parallel strategy for each layer in the original graph is determined.


Let $C$ be the maximum number of potential configurations for parallelism in a layer, $|E|$ denote the number of nodes and edges in $G$, respectively, and $|E'|$ be the number of nodes in the simplified graph $G'$.
The time complexity of performing node and edge eliminations is $O(|E|\cdot C^3)$, and for identifying the optimal partitioning solution, it is $O(|E'|^2)$. In addition, HyPAD explores all possible parallelism strategies for $G'$, resulting in a time complexity of $O(|E'| \cdot {C}^{|E'|})$. Therefore, the total complexity is $O(|E|\cdot C^3+|E'|^2+K \cdot {C}^{|E'|})$. Note that $C$ and $|E'|$ are relatively small values, typically less than 10 in our experiments.

\subsection{Communication Optimization Module} 
\label{sec:COM}

Due to the stateless nature of serverless computing, external storage is required for data exchange between slices, which leads to an increase in communication time. Therefore, although MPE can improve resource utilization efficiency, it may increase additional latency. We design a communication optimization module (COM) in MOPAR to systematically offset the increase in latency from two aspects. 
Firstly, to reduce the amount of data transferred between slices, COM leverages the \textbf{auto-encoder (AE)} to encode the tensor output into a smaller dimension, reducing the communication overhead.
Secondly, the \textbf{share-memory} mechanism is used in COM to speed up inter-slice data exchange.

\begin{figure}[tb]
\begin{center}
\includegraphics[width=\linewidth,scale=1.00]{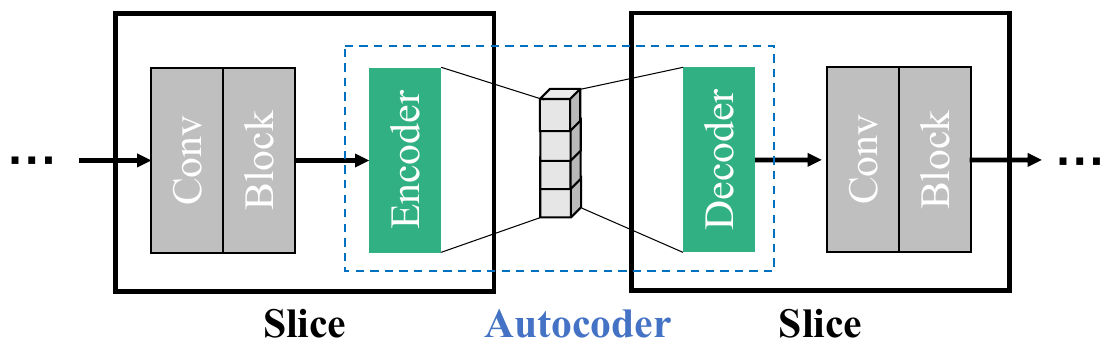}
    \caption{Autoencoder for inter-slice data transfer}
    \label{fig:encoder}
\end{center}
\vspace{-7mm}
\end{figure}

\noindent
\textbf{AE for Inter-slice Data Transmission.} As demonstrated in Section \ref{sec:obser}, data compression has a significant impact on reducing communication latency, where AE exhibits enhanced capacity to apprehend data patterns with varying distributions. COM uses a general pre-trained AE to facilitate data coding between slices \cite{quiltnet}, as depicted in Fig. \ref{fig:encoder}. The AE consists of an encoder and a decoder, which maps (compresses) the original tensor $T_o$ into a smaller dimension, denoted as $T_e$, and restores it to the original size, respectively \cite{quiltnet}. The compression ratio is denoted as $R = \frac{|T_o|}{|T_e|}$. 
Specifically, the AE adopts a 2D convolutional structure. Each encoder and decoder use one layer respectively. The AE takes the DLIS model structure as input and inserts the encoder and decoder in the two consecutive slices respectively at each split point. 

To ensure the generality of the AE, COM employs data augmentation techniques to create different variants of input data, such as image skewing and cropping, so that the AE can better learn common patterns.
Park et al. verified that the AE generates negligible accuracy losses \cite{quiltnet}. On average, the accuracy using the AE only decreases by 0.05\% compared to the approach without data compression. Importantly, it should be emphasized that the compression ratio can be adjusted to achieve higher levels of accuracy.

\begin{figure}[tb]
\begin{center}
\includegraphics[scale=0.42]{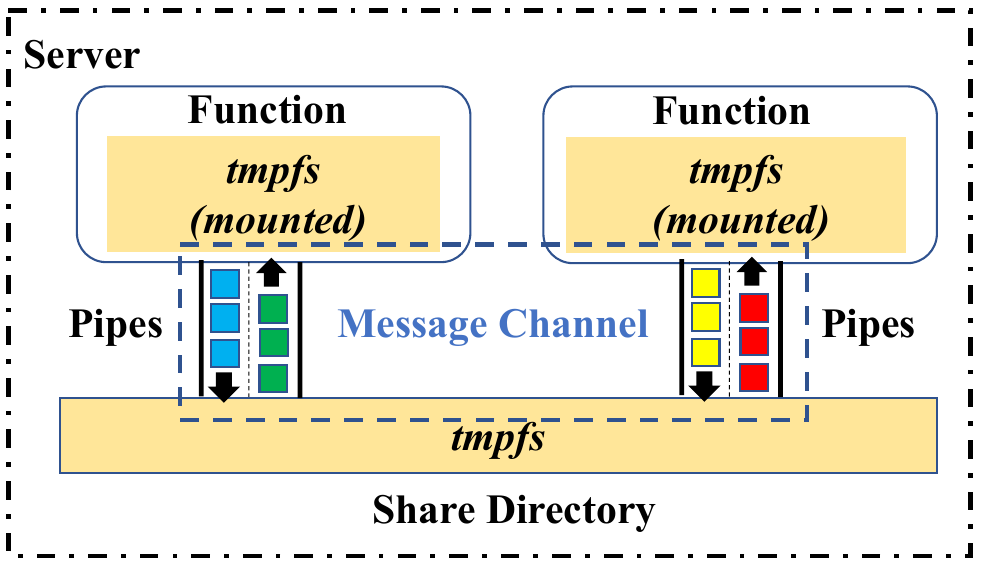}
    \vspace{-0.2cm}
    \caption{Share-memory for slice communication}
    \label{fig:share_memory}
\end{center}
\vspace{-5mm}
\end{figure}

\noindent
\textbf{Share-memory for Slice Communication.}
The COM employs the share-memory mechanism to reduce data copying between slices, leading to faster and more efficient communication. 
Share-memory and message channels are integral to enabling fast slice communication, as shown in Fig. \ref{fig:share_memory}. 
First, share-memory is utilized to obtain inputs and generate outputs for slices \cite{Nightcore}. The design of share-memory in COM is inspired by the scheme in Linux's System V\footnote{https://en.wikipedia.org/wiki/UNIX\_System\_V}, which is implemented by mounting the \textbf{\emph{tmpfs}} directory. Specifically, a shared \textbf{\emph{tmpfs}} directory is mounted in the two consecutive slices at each split point. 
Second, data read and write synchronization on the shared \textbf{\emph{tmpfs}} directory between slices is achieved through a message channel, which creates a pair of Linux pipes to facilitate a full-duplex connection.

COM also provides function templates that enable multiple slices of a DLIS to be deployed on the same server. The function templates add an \emph{affinity} field which comprises the \emph{preferredDuringSchedulingIgnored} policy to prioritize scheduling the slices on the same server. In cases where all candidate servers cannot meet this deployment condition, slice scheduling follows the random policy.

\section{Experiments}
\label{sec:eva}

\subsection{Performance Evaluation on CPUs}

\subsubsection{Setup}

\begin{figure}[tb]
    \centering
    \includegraphics[width=0.9\linewidth]{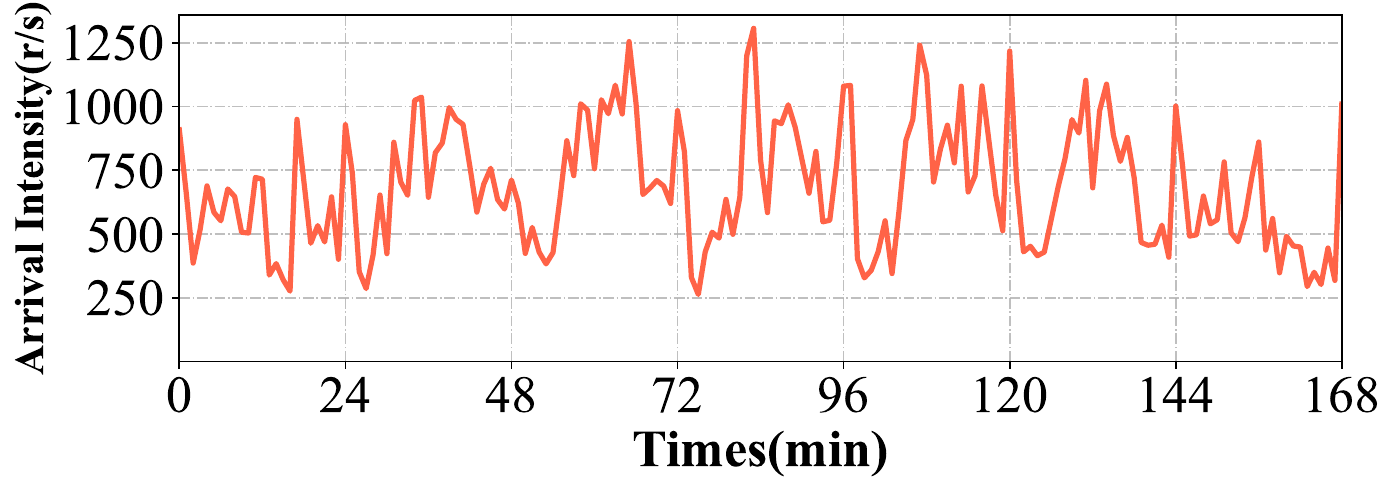}
      \vspace{-0.3cm}
\caption{The trace of generated service requests}
\vspace{-0.4cm}
\label{fig:trace}
\end{figure}

\textbf{Testbed}:
We verify the effectiveness of MOPAR on two serverless platforms: the open-source OpenFaaS\footnote{https://www.openfaas.com/} and the widely used platform AWS Lambda. 
We evaluate MOPAR using two DL frameworks, namely PyTorch \cite{2019pytorch} and MXNet \cite{2015mxnet}, and bundle the DL library required for DLISs in a container. 
Table \ref{table:Experimental configuration} presents the configurations of 14 servers on which we setup OpenFaaS. 
Motivated by prior work \cite{weng2022mlaas}, we generate the workload of DLIS requests according to the diurnal service submissions from PAI, as shown in Fig. \ref{fig:trace}. The arrival rate of submissions varies from 250 to 1250 requests/second, with data size from 100KB to 100MB.
\begin{scriptsize}
\begin{table}[tb]
  \centering
  \caption{Hardware configuration of servers}
   \resizebox{\linewidth}{!}{
  \begin{tabular}{| c| c |c| c|} 
    \hline
    \textbf{Component} & \textbf{Specification} &\textbf{Component} & \textbf{Specification}\\
    \hline
    CPU Model & Intel(R) Xeon(R) 8269CY & Shared LLC Size & 36MB \\
    Number of sockets & 2 & Memory Capacity & 187GB \\
    Processor Base Freq. & 3.2 GHz & Operating System & Ubuntu 18.04\\
    CPU Threads & 104 (52 physical cores) & SSD Capacity & 2.0TB \\\hline   
  \end{tabular}
  }
 \label{table:Experimental configuration}
 \vspace{-2mm}
\end{table}
\end{scriptsize}


\begin{figure*}[htb]
    \centering
    \includegraphics[scale=0.5]{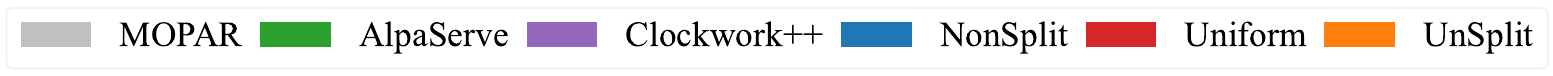}

    \vspace{-5mm}  
   \subfloat[Normalized memory utilization]{
    \includegraphics[scale=0.25]{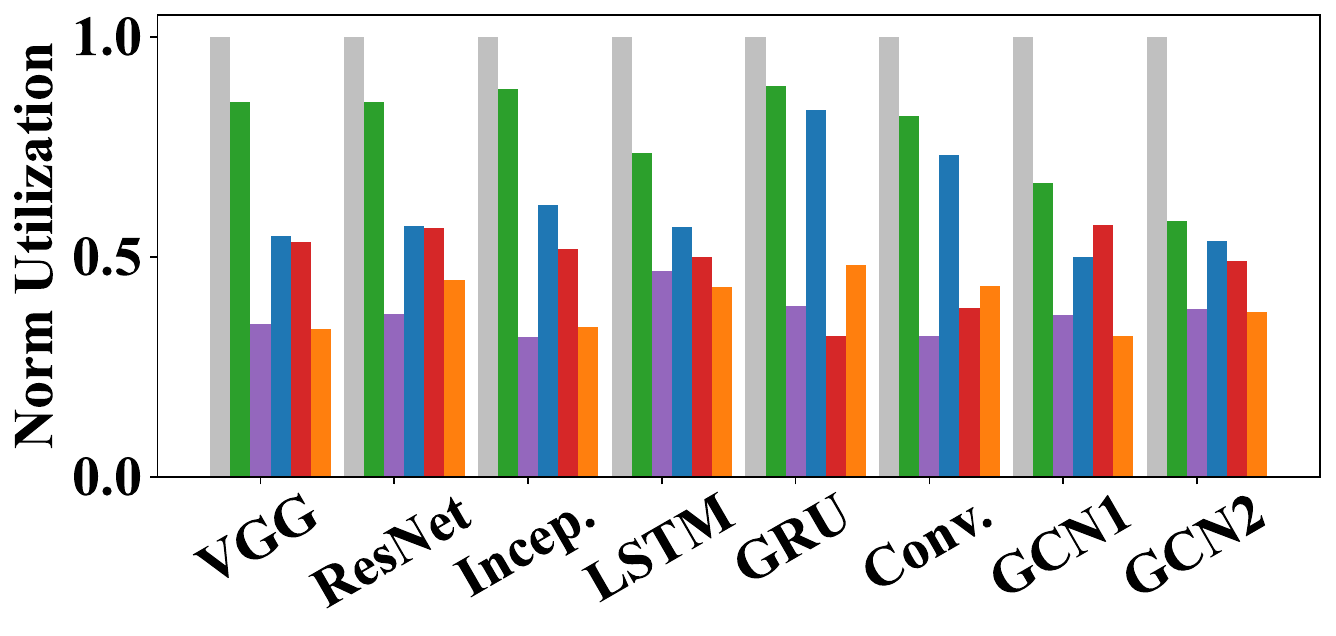}
        \label{fig:memory_foot}
  }
  \subfloat[Latency in OpenFaaS]{
    \includegraphics[scale=0.25]{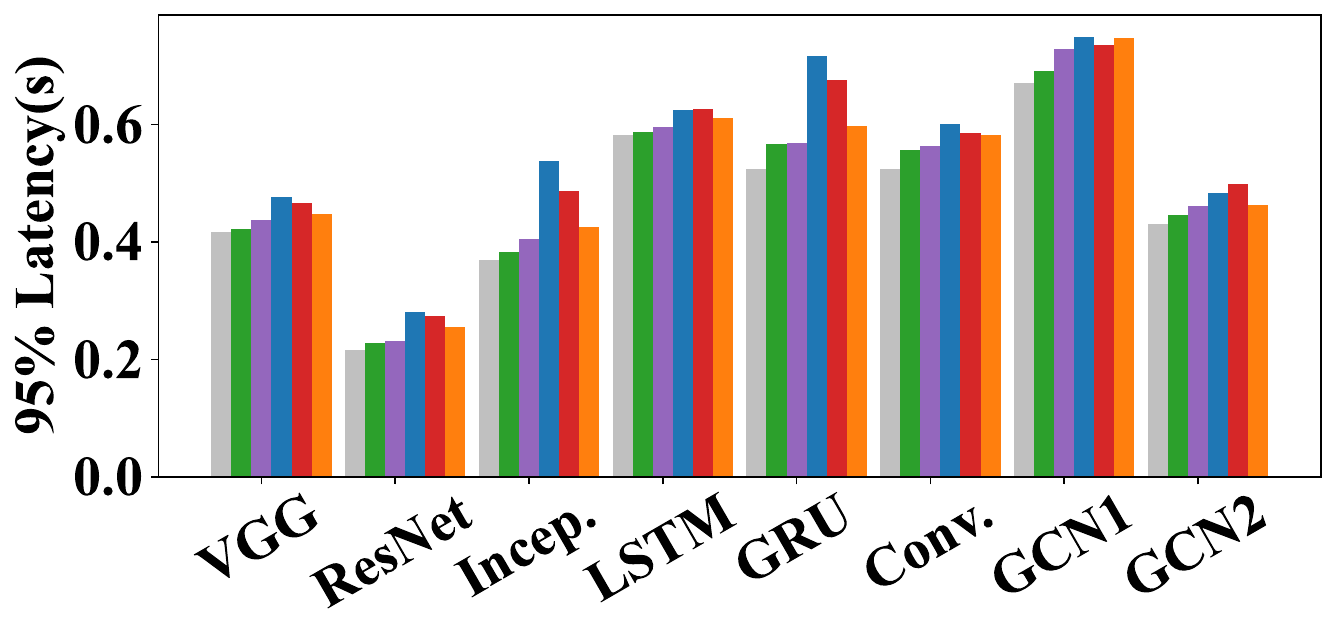}
        \label{latency}
  }
  \subfloat[Latency in Lambda]{
    \includegraphics[scale=0.25]{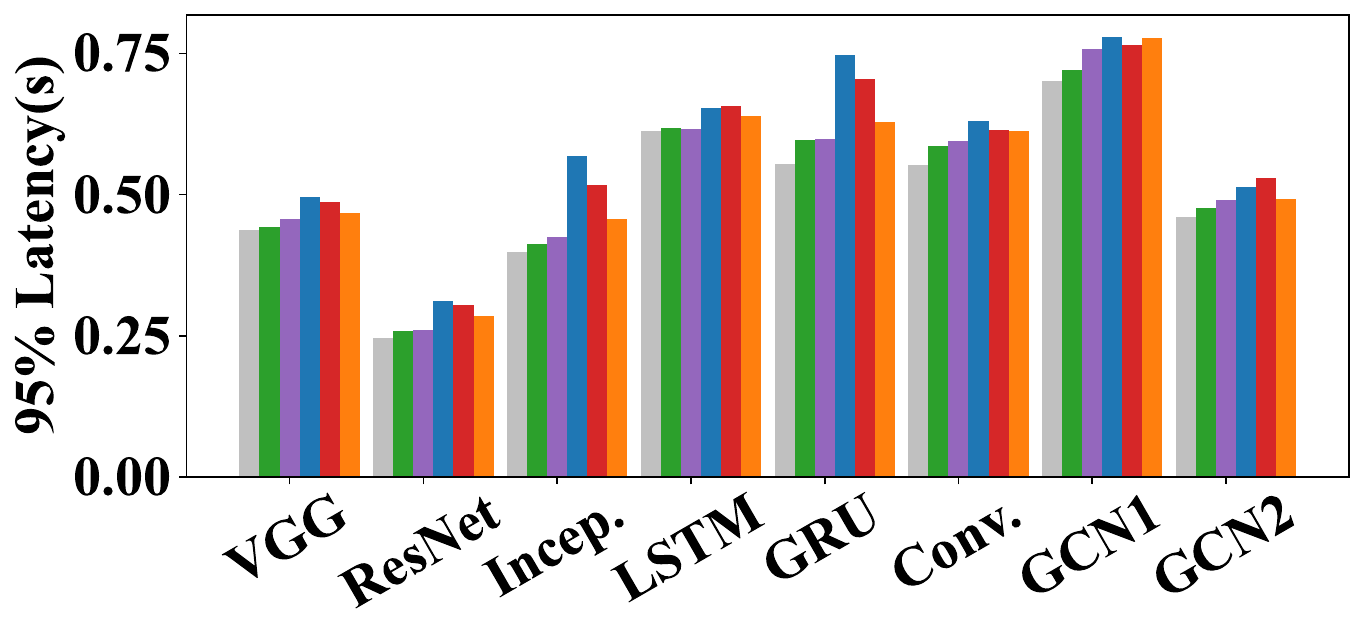}
        \label{fig:aws}
  } 
\caption{ The normalized memory utilization and 95th percentile latency of DLISs with different baselines}
\vspace{-5mm}
\end{figure*}

\noindent
\textbf{Benchmarks}: We test four types of twelve DL models as benchmarks to study the partition effect of MOPAR, including four CNNs, two RNNs and two GCNs. (1) CNNs are commonly used for computer vision, utilizing convolutional layers to automatically learn hierarchical features from input data. We use VGG \cite{VGG16}, ResNet \cite{resnet}, Inception \cite{Inception}, and ConvNeXt \cite{convNeXt}. (2) RNNs are designed to process sequential data using recurrent layers to retain information from previous time steps. We use LSTM+CNN \cite{LSTM-2365} and GRU+CNN \cite{gru}. (3) GCNs are designed to operate on graph-structured data, effectively capturing relationships and dependencies between nodes in a graph. We use two GCNs based on the previous work \cite{GCN}. (4) We analyze the effect of MOPAR on four transformer-based DLISs, namely BERT-1.3B \cite{bert}, BERT-3.0B \cite{bert}, DisBERT \cite{yu2022learning}, and Transformer-2.6B \cite{si2022inception}. 

\noindent
\textbf{Baselines}: We conduct a comparative analysis of MOPAR against five baselines, which include four partition algorithms and one unsplit method. (1) \emph{AlpaServe}\cite{li2023alpaserve} is a system that incorporates model parallelism and placement techniques, employing a dynamic programming algorithm to partition DLIS into several distinct slices, all with the ultimate objective of decreasing inference latency. 
(2) \emph{Clockwork++} is an improved version of the advanced model serving system known as Clockwork \cite{clockwork}. Clockwork++ integrates a placement algorithm that prioritizes selecting the same server which facilitates the exchange of models at the interface between two consecutive trace windows, in accordance with Clockwork's replacement strategy.
(3) \emph{Non-contiguous Split (NonSplit)} \cite{tarnawski2020efficient} focuses on optimizing the latency of DLISs, which employs an integer programming approach to exploit the parallelism of non-contiguous subgraphs within DLISs.
(4) \emph{Uniform} is an even partitioning strategy, which divides a DLIS into the same number of slices as MOPAR, so each slice has the same number of layers.
(5) \emph{Unsplit} is the default strategy, which deploys a DLIS as a function without partitioning.







%


\subsubsection{Comparison Results}

We focus our efforts on validating the impact of MOPAR in terms of latency, resource utilization, and costs.

\noindent
\textbf{Memory Utilization:}
We conducted a comparative analysis of the six approaches to evaluate their impact on memory utilization. Fig. \ref{fig:memory_foot} shows the average memory utilization of different approaches in Lambda and OpenFaaS normalized over MOPAR. The results indicate that MOPAR notably improves memory utilization for the three types of eight non-transformer-based DLISs, surpassing the performance of the five baselines. Compared to AlpaServe, NonSplit, Uniform, Clockwork++, and Unsplit,
MOPAR demonstrates superior performance with respective improvements of 15.99\%, 23.43\%, 27.82\%, 34.2\%, and 36.65\%.
MOPAR groups layers with similar resource requirements into a slice, thereby ensuring the maximal utilization of resources within each slice.
By contrast, AlpaServe and NonSplit prioritize the optimization of slicing latency, disregarding resource utilization within each slice.
Furthermore, the five partitioning approaches outperform the unsplit method on the whole.

\begin{figure*}[tb]
    \centering
    \includegraphics[scale=0.5]{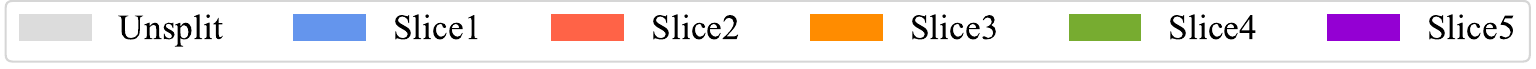}

    \vspace{-4mm}
  \subfloat[ResNet50]{
    \includegraphics[scale=0.25]{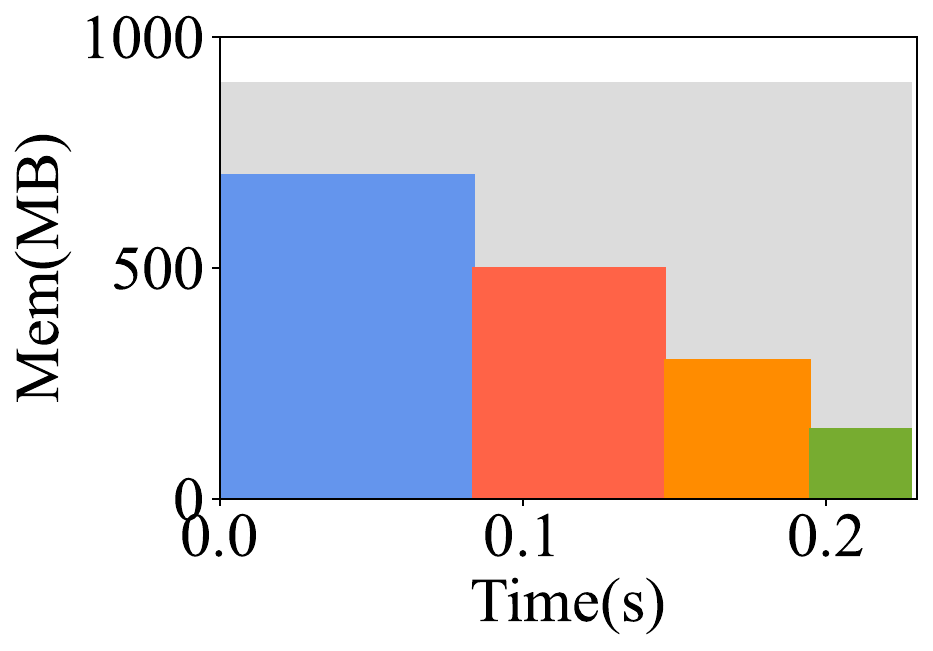}
        \label{fig:3/Resnet_usg}
  }
  \subfloat[VGG]{
    \includegraphics[scale=0.25]{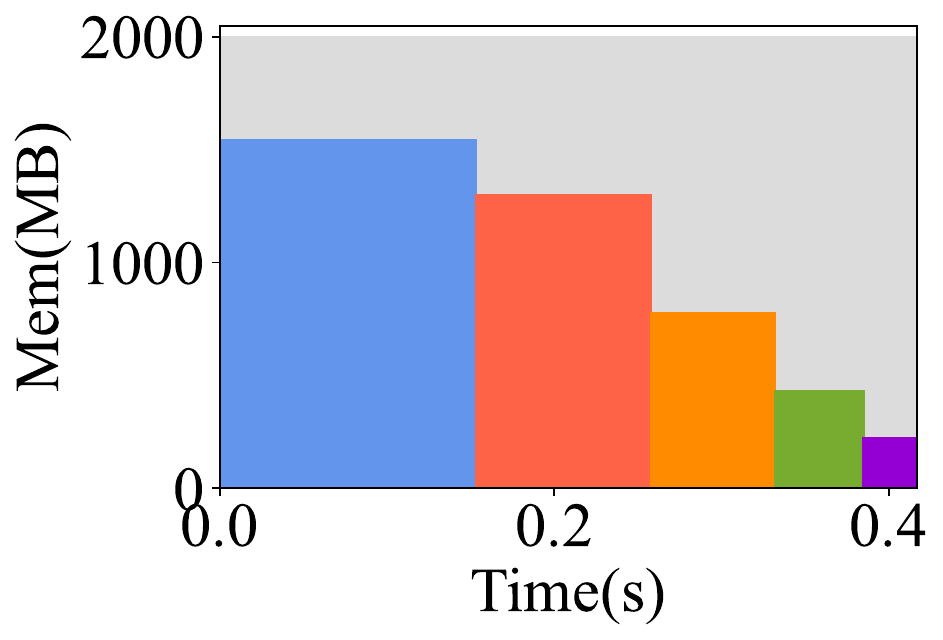}
        \label{fig:3/vgg_usg}
  }
  \subfloat[Inception]{
    \includegraphics[scale=0.25]{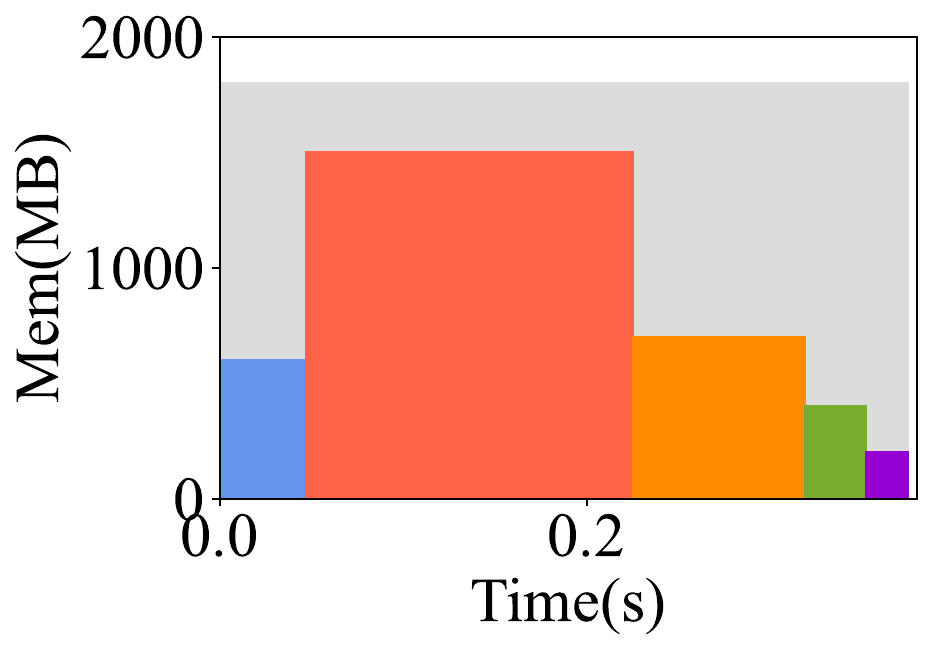}
        \label{fig:3/rnn6_usg}
  }
  \subfloat[ConNeXt]{
    \includegraphics[scale=0.25]{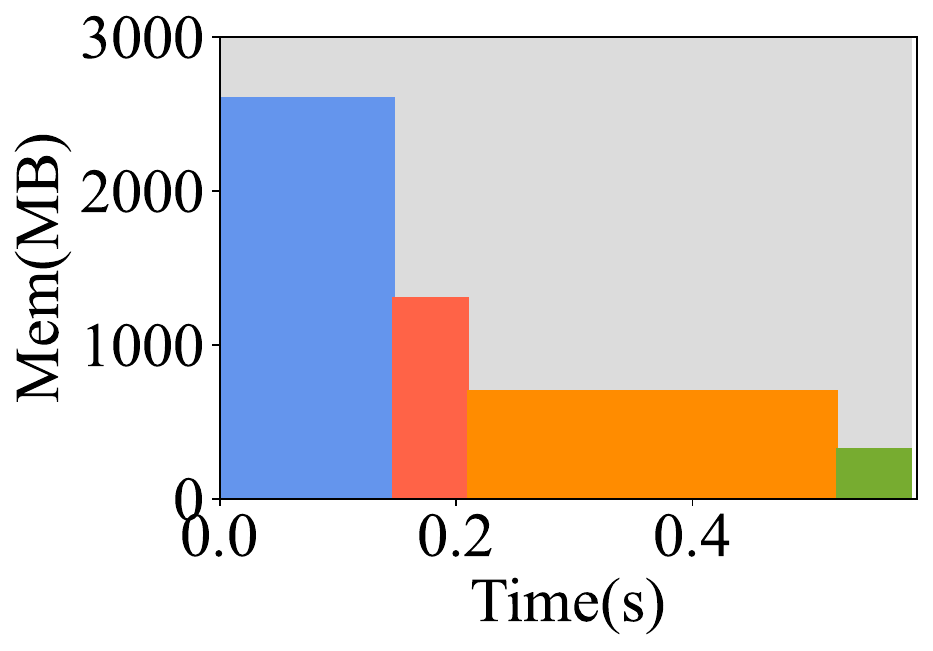}
        \label{fig:3/convnet_usg}
    
    }
  \vspace{-0.3cm}
  
  \subfloat[LSTM+CNN]{
    \includegraphics[scale=0.25]{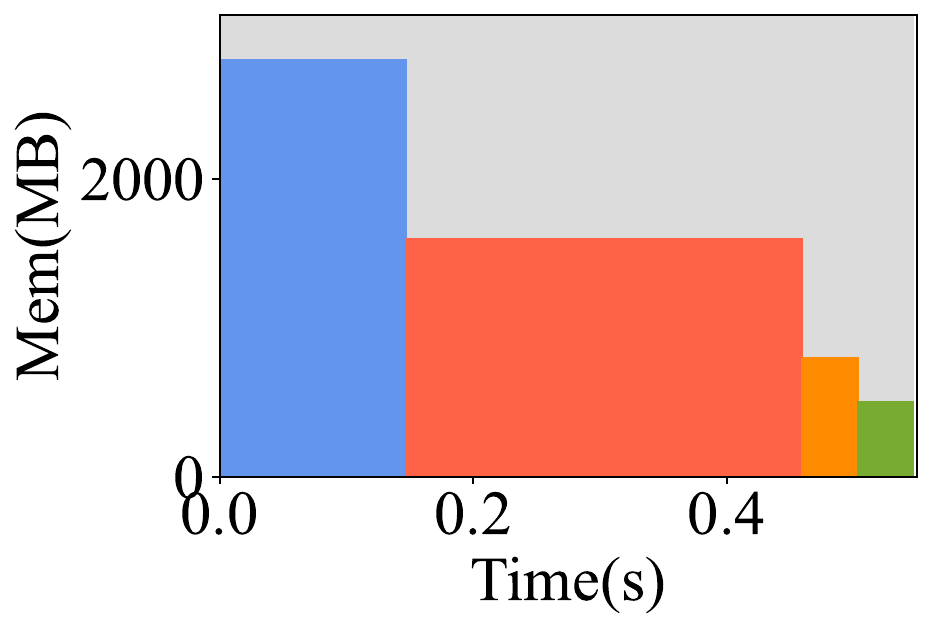}
        \label{fig:3/rnn3_usg}
  }
  \subfloat[GRU+CNN]{
    \includegraphics[scale=0.25]{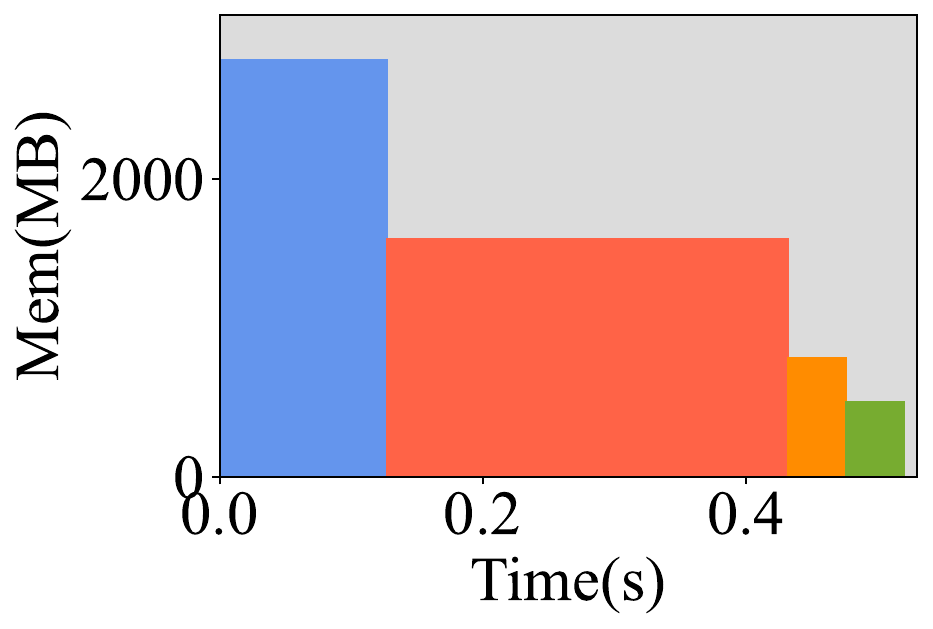}
        \label{fig:3/gcn_usg}
  }
  \subfloat[GCN1]{
    \label{fig:3/yolo_usg}
    \includegraphics[scale=0.25]{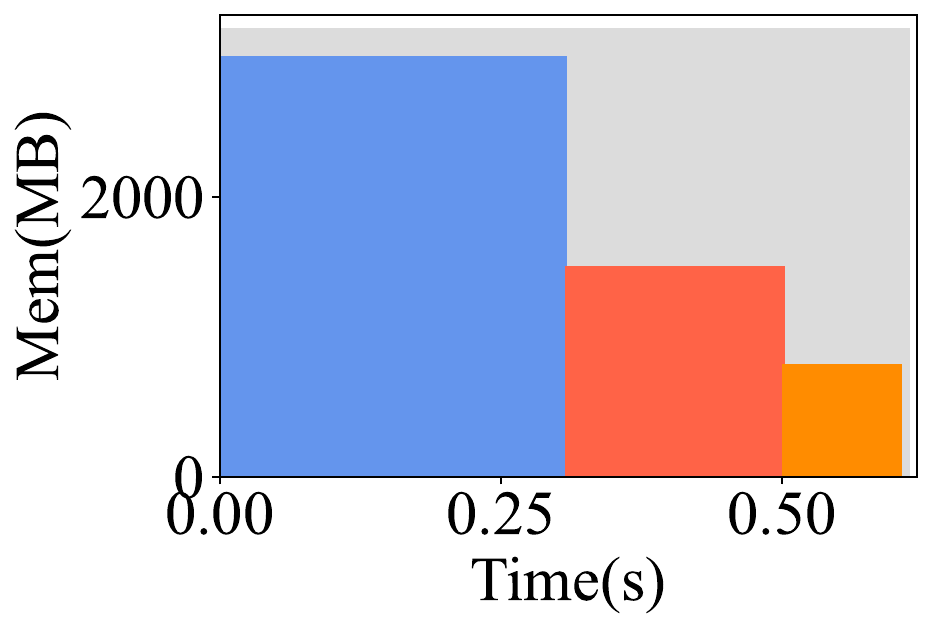}
    }
  \subfloat[GCN2]{
    \includegraphics[scale=0.25]{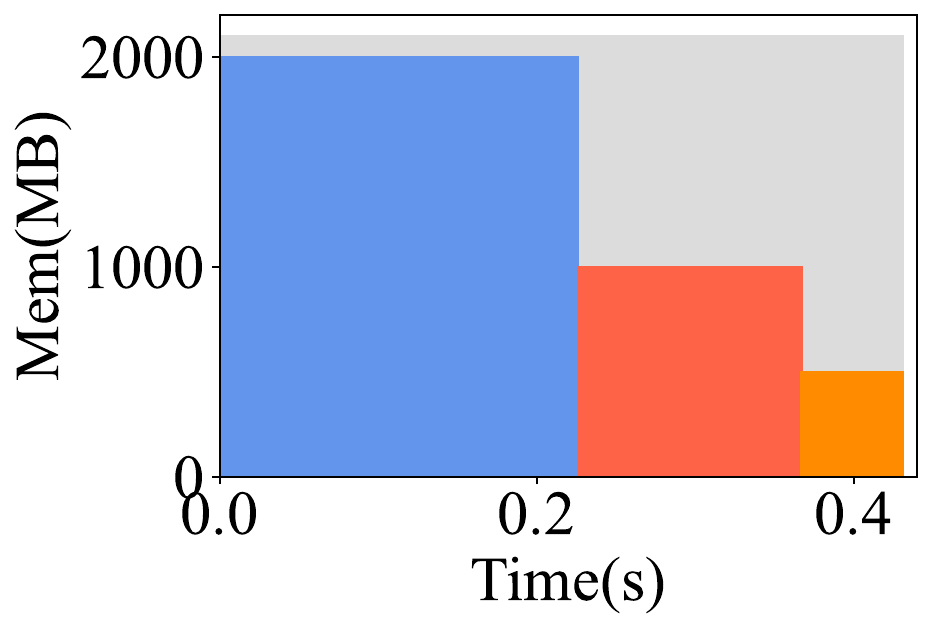}
        \label{fig:3/unet_usg}
  } 
\caption{The memory footprint of MOPAR and Unsplit}
\label{fig:memory_footprint}
\vspace{-4mm}
\end{figure*}

\noindent
\textbf{Latency:}
Fig. \ref{latency} provides a comparison of the 95th percentile latency across the eight DLISs, considering various workloads for the six tested methods in OpenFaaS. We observe that MOPAR performs significantly better than the five baselines.
Specifically, compared to Unsplit, MOPAR achieves an average latency reduction of 5.47\%, 6.71\%, and 9.31\% for CNNs, RNNs, and GCNs, respectively. This notable improvement can be attributed to the systematic optimization of COM. Furthermore, note that the influence of SP and MPE on latency is negligible, accounting for only 0.074\%.
Similarly, compared to the Unsplit method in Lambda, MOPAR demonstrates average latency reductions of 3.82\%, 5.13\%, and 7.01\% for CNNs, RNNs, and GCNs, respectively, as depicted in Fig. \ref{fig:aws}. 
The relatively lower performance in Lambda is due to the absence of support for shared-memory mechanisms. Despite the improvement in communication bandwidth, data transmission between functions still necessitates the utilization of AWS ElastiCache \cite{cache}. Overall, MOPAR exhibits 2.58\% higher latency in Lambda compared to OpenFaaS.

Furthermore, in comparison to the other four partitioning baselines, MOPAR yields superior results, with an average latency reduction of 19.19\%. Both MOPAR and AlpaServe exhibit lower latency than Unsplit due to their implementation of parallel strategies to enhance computing efficiency. Additionally, MOPAR further mitigates latency by utilizing data compression and learning-driven methods.

\noindent
\textbf{Costs:}
To validate the economics of MOPAR, we compute the memory consumption cost per DLIS request. The pricing model of commercial serverless platforms is based on factors such as the amount of memory allocated to the function, runtime, and data transfer volume.
In accordance with Lambda's pricing\footnote{https://aws.amazon.com/lambda/}, we set a price per 1 millisecond associated with different memory sizes for MOPAR. The minimum configuration is 128MB, and the price per GB per second is $\$1.667\times 10^{-5}$ on average. Table \ref{table:computation cost comparison} shows that MOPAR reduces the cost by about $5.6 \times$ compared to that of Unsplit in OpenFaaS. 
In Lambda, we simulate the COM mechanism in MOPAR by increasing the network transmission rate. Overall, MOPAR reduces cost by $2.58\times$ compared to the default Unsplit scheme in Lambda.
The primary source of cost savings stems from three key aspects. Firstly, there is a reduction in resource utilization owing to the MPE in MOPAR. Secondly, MOPAR expedites the execution time of DLISs through the implementation of a parallelization strategy. Lastly, data compression contributes to reducing communication costs.

\subsubsection{Transformer-based DLISs}
\label{sec:transformer}
\begin{figure}[tb]
    \centering
  \vspace{-4mm}
  \subfloat[Resource consumption]{
    \includegraphics[width = 0.54 \linewidth]{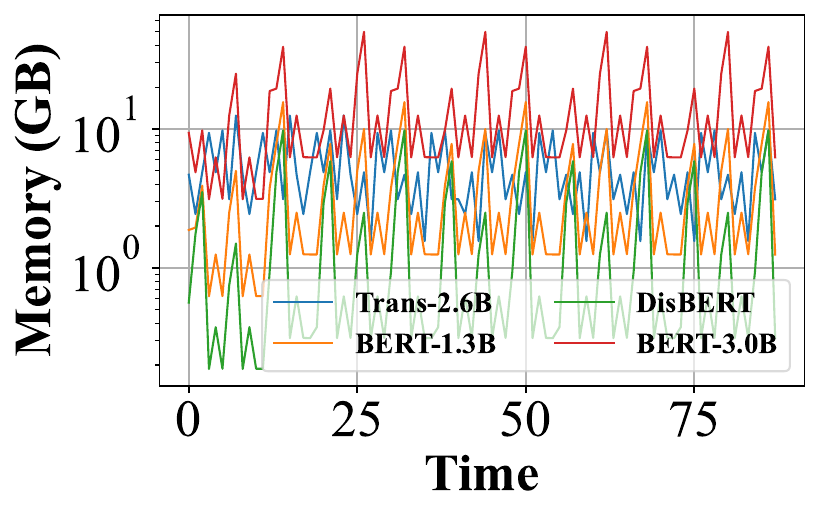}
        \label{fig:tra_ope}
  }
  \subfloat[Latency]{
    \includegraphics[width= 0.4 \linewidth]{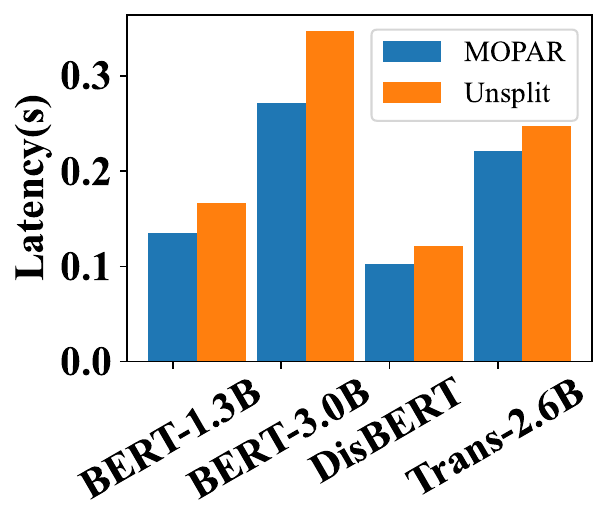}
        \label{fig:tra_latency}
  }

\caption{Resource consumption and latency of Transformer-based DLISs}
\label{transformer}
\vspace{-4mm}
\end{figure}

Transformer-based DLISs have gained significant popularity, particularly with the emergence of ChatGPT \cite{chatgpt}. Due to the inclusion of multiple self-attention layers, Transformer-based DLISs necessitate the execution of a weighted mean operation on each element within the input sequence to capture inter-element dependencies. Consequently, the consumption of resources of Transformers exhibits fluctuations, as depicted in Fig. \ref{fig:tra_ope}.
Although the resource fluctuations of Transformers hinder the enhancement of resource utilization through MP, MOPAR enables a reduction in the latency for Transformer-based DLISs by utilizing the horizontal parallelization scheme, as illustrated in Fig. \ref{fig:tra_latency}. MOPAR reduces latency for all four Transformers by an average of 16.63\%.

\begin{table}[tb]
  \centering
  \caption{Cost comparison of methods}
 \vspace{-2mm}
    \resizebox{\linewidth}{!}{
  \begin{tabular}{c c c c c c c} 
    \hline
      & \textbf{} & \textbf{Unsplit} & \textbf{NonSplit} &\textbf{Clockwork++} & \textbf{AlpaServe} &\textbf{MOPAR}\\
    \hline
   \multirow{2}{*}{\textbf{OpenFaaS}} &Memory per 100RPS  & 8.291 & 7.128 & 7.385 &  6.437 & 3.531 \\
    &Cost per request ($10^{-6}$\$) & 29.2 & 29.6 & 30.3  & 20.6 & 3.68 \\\hline
    \multirow{2}{*}{\textbf{Lambda}}&Memory per 100RPS  & 8.311 & 7.245 & 7.694 & 6.612 & 3.661 \\
    &Cost per request ($10^{-6}$\$) & 32.3 & 34.1 & 37.5  & 27.6 & 10.7 \\\hline
  \end{tabular}
  }
  \label{table:computation cost comparison}
  \vspace{-3mm}
\end{table}

\subsection{Evaluation with Large Language Models (LLM) on GPUs}
\label{sec: chatongpu}
The utilization of GPUs in DLISs holds immense potential in enhancing computational performance and efficiency, especially with the development of Chat Generative Language Models (GLMs), such as ChatGPTs \cite{chatgpt} and ChatGLMs \cite{du2022glm}. By exploiting the parallel processing capabilities of GPUs, LLMs can effectively handle the complex and resource-intensive tasks involved in natural language processing. This integration of GPUs with LLMs enables the acceleration of various computational operations, such as text generation, word embeddings, and attention mechanisms, resulting in significant improvements in the overall speed and responsiveness of LLMs. MOPAR remains highly effective for LLM inference on GPUs.

\subsubsection{Setup} 
\textbf{Testbed:}
We evaluate the performance of MOPAR on two 2-A40 GPU-enabled servers focusing on the efficiency of LLM inference generation using ChatGLM, an encoder-decoder converter model\cite{du2022glm}. Each GPU possesses a memory capacity of 48GB, with NVLink enabling a swift data transfer rate of 112.5 GB/s between GPUs. The network employed among servers is the Mellanox Technologies MT2894 Family[ConnextX-6 Lx] \cite{RDMA1}, which offers a speed of 50GB/s. We utilize the NVIDIA Container Toolkit\cite{nvidiatoolkit} to facilitate container access to the GPU.

\noindent
\textbf{Benchmarks:}
We assess the character generation rates across five ChatGLMs of varying sizes as benchmarks, including two ChatGLMs, two ChatGLM2s, and one ChatGLM3. ChatGLMs serve as an open-source conversational language model that enables bilingual question answering in both Chinese and English. These models are built upon the GLM architecture, featuring an extensive composition of 28 layers and a total parameter count of 6.2 and 12 billion. ChatGLMs utilize the official implementation with PyTorch 2.0.1, with the accuracy being represented by $bfloat16$. 

\noindent
\textbf{Baselines:}
We perform a comparative examination of MOPAR in relation to two baselines: NonSplit and Default. 1) The NonSplit strategy entails deploying the entire model on a GPU. 2) The Default strategy employed by PyTorch evenly distributes the model's weights into multiple parts, with each part assigned to a GPU.
Additionally, we use zero-shot CoT (Chain-of-Thought) \cite{kojima2022large} for testing in Massive Multitask Language Understanding (MMLU)\cite{MMLU}, as well as C-Eval \cite{huang2024c}. Drawing from \cite{du2022glm}, we use characters per second (char/s) and accuracy to measure the tested models.

\subsubsection{Comparison Results} Our findings focus on the verification of the influence of MOPAR in terms of inference speed, accuracy, and communication costs.

\noindent
\textbf{Inference Speed \& Accuracy}:
We separately verify the inference speed and accuracy of five ChatGLMs under different strategies.
First, MOPAR achieves a notable improvement of 6.48\% in inference speed while experiencing a mere 0.03\% decrease in accuracy. These experimental outcomes are presented in Table \ref{table:Inference Speed} and Table \ref{table:Inference Statistics}, which shows that MOPAR achieves significant inference speed improvements with 2 and 4 GPUs. Specifically, compared to Default, MOPAR's inference speed increases by 6.32\% with 2 GPUs and 6.64\% with 4 GPUs on average.
Second, MOPAR demonstrates significant improvement compared to NonSplit, with 70.62\% faster inference. This significant improvement can be attributed to the full utilization of the resources of multiple GPUs, resulting in increased inference speed. Additionally, when the model size becomes larger (e.g., ChatGLM2-12B), the NonSplit strategy will fail due to the GPU's out-of-memory (OOM) errors. 

\noindent
\textbf{Communication Cost}:
We further analyze the reasons for the improved MOPAR effect and conduct experimental comparison and verification.
We collect the statistics of MOPAR and Default in terms of communication time, as presented in Table \ref{table:Inference Statistics}. 
In comparison to Default, MOPAR achieves an 18.96\% improvement in communication efficiency. 
Both the Default and MOPAR strategies execute DLISs on multiple GPUs and multiple servers (e.g., ChatGLM3-12B, etc.), resulting in network communication between servers becoming a performance bottleneck. MOPAR uses the AE method to improve communication efficiency and speed up generation while ensuring that data loss is negligible. 
In addition, GPUs possess a comprehensive parallelism mechanism, employing a large-scale parallel architecture and featuring an instruction set specifically designed for parallel computing. As a result, MOPAR can effectively leverage the inherent parallelism characteristics offered by GPUs.

\begin{table}[tb]
  \centering
  \caption{Inference speed (char/s) of ChatGLMs with different strategies on GPUs}
  \resizebox{\linewidth}{!}{
  \begin{tabular}{c c c c c c} 
    \hline
    & & \multicolumn{2}{c}{2 GPUs} & \multicolumn{2}{c}{4 GPUs} \\
       & \textbf{NonSplit} & \textbf{Default} &\textbf{MOPAR} & \textbf{Default} & \textbf{MOPAR}\\ [0.5ex]
    \hline
    ChatGLM-6B & 20.04 & 32.47 & 34.71 & 59.49& 62.52  \\
    ChatGLM2-6B& 28.16 & 46.01 & 49.11 & 82.53&87.07 \\
    ChatGLM3-6B  & 28.87 & 46.77 & 49.26 & 83.89& 89.73\\
    ChatGLM2-12B & \multicolumn{1}{c}{--} & 27.31 & 29.24 & 39.29& 42.12\\
    ChatGLM3-12B & \multicolumn{1}{c}{--} & 27.98 & 29.98 & 42.26& 45.17\\\hline
  \end{tabular}
  }
  \label{table:Inference Speed}
\end{table}

\begin{table}[tb]
  \centering
  \caption{Performance comparison with baselines}
  \resizebox{\linewidth}{!}{
  \begin{tabular}{c c c c c} 
    \hline
    & & \textbf{NonSplit}&\textbf{Default} &\textbf{MOPAR}\\ [0.5ex]
    \hline
    \multirow{2}{*}{ChatGLM-6B} &Accuracy (\%) & 40.63 & 40.62 & 40.59 \\
    &Communication(ms)& 0 & 56 & 43 \\\hline
    \multirow{2}{*}{ChatGLM2-6B} &Accuracy (\%)& 45.46 & 45.44 & 45.42 \\
    &Communication(ms) & 0 & 42 & 33 \\
    \hline
    \multirow{2}{*}{ChatGLM2-12B} &Accuracy (\%)& \multicolumn{1}{c}{--} & 52.12 & 52.09 \\
    &Communication(ms) & \multicolumn{1}{c}{--} & 96 & 79 \\\hline
    \multirow{2}{*}{ChatGLM3-12B} &Accuracy (\%)& \multicolumn{1}{c}{--} & 61.37 & 61.33 \\
    &Communication(ms) & \multicolumn{1}{c}{--} & 89 & 77 \\\hline
  \end{tabular}
  }
  \label{table:Inference Statistics}
\end{table}

\subsubsection{Discussion}
Recently, the spatial parallelization of GPUs, which involves deploying and executing multiple DLISs concurrently on a GPU, lacks adequate support in serverless computing environments. Serverless computing platforms primarily cater to CPU computing tasks and data processing to achieve fine-grained and efficient resource management. While some cloud service providers have made efforts to integrate serverless computing with GPU capabilities, e.g., AWS Lambda's GPU instances\footnote{https://aws.amazon.com/lambda/}, it is important to note that they only enable time division of GPUs and do not yet possess the capability to facilitate spatial sharing of GPUs. 

We believe that in-depth exploration of GPU virtualization technology will contribute to the advancement of GPUs, enabling their utilization in more complex scenarios similar to CPUs. For example, the NVIDIA A100 supports the execution of multiple instances on a single GPU.
Furthermore, LLMs consist of numerous parameters and exhibit substantial size. However, existing studies \cite{ma2024llm} have demonstrated that the same effect can be achieved by compressing LLMs, thereby significantly reducing the size of LLMs. Consequently, there will undoubtedly be situations in the future where multiple LLMs will be deployed on a single GPU. 
Based on the aforementioned points, the significance of MOPAR becomes more evident. We will continue to explore the optimization of inference services for large models on GPUs.

\begin{figure*}[tb]
    \centering
   \subfloat[MPE for MOPAR]{
    \includegraphics[scale=0.25]{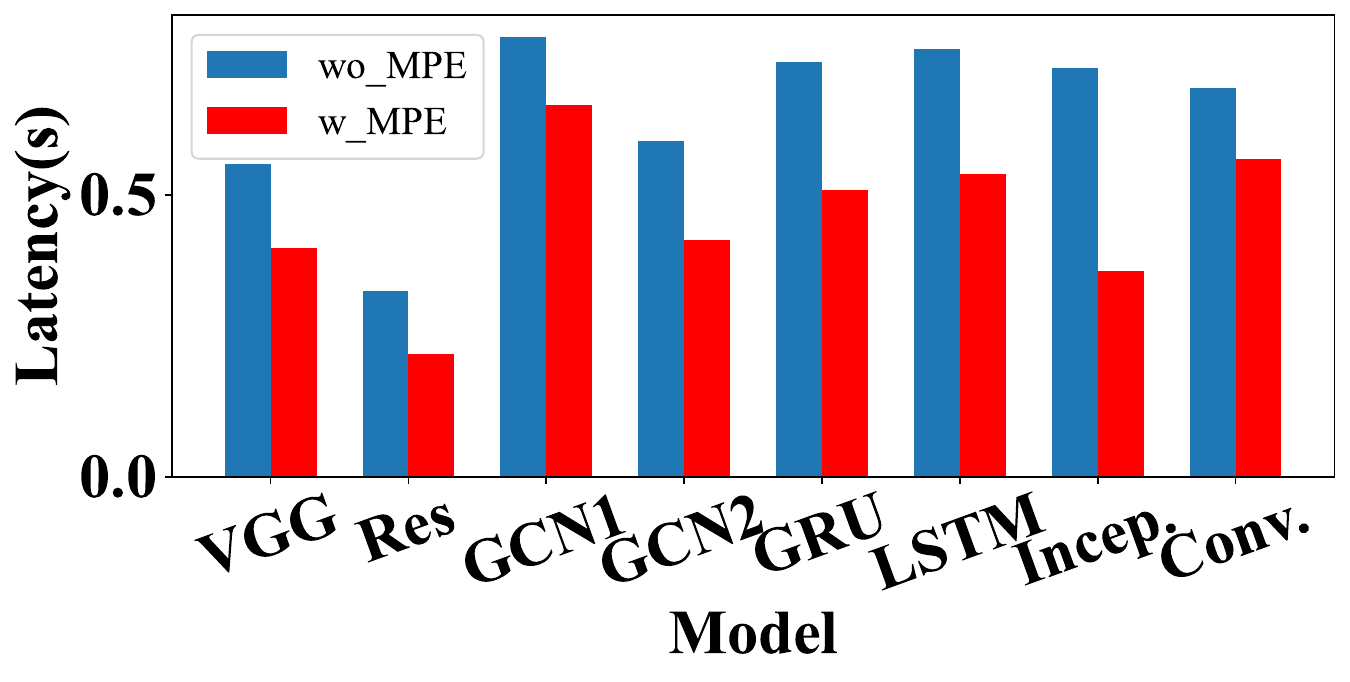}
        \label{fig:3/parallel}
  }
  \subfloat[Share-memory for MOPAR]{
    \includegraphics[scale=0.25]{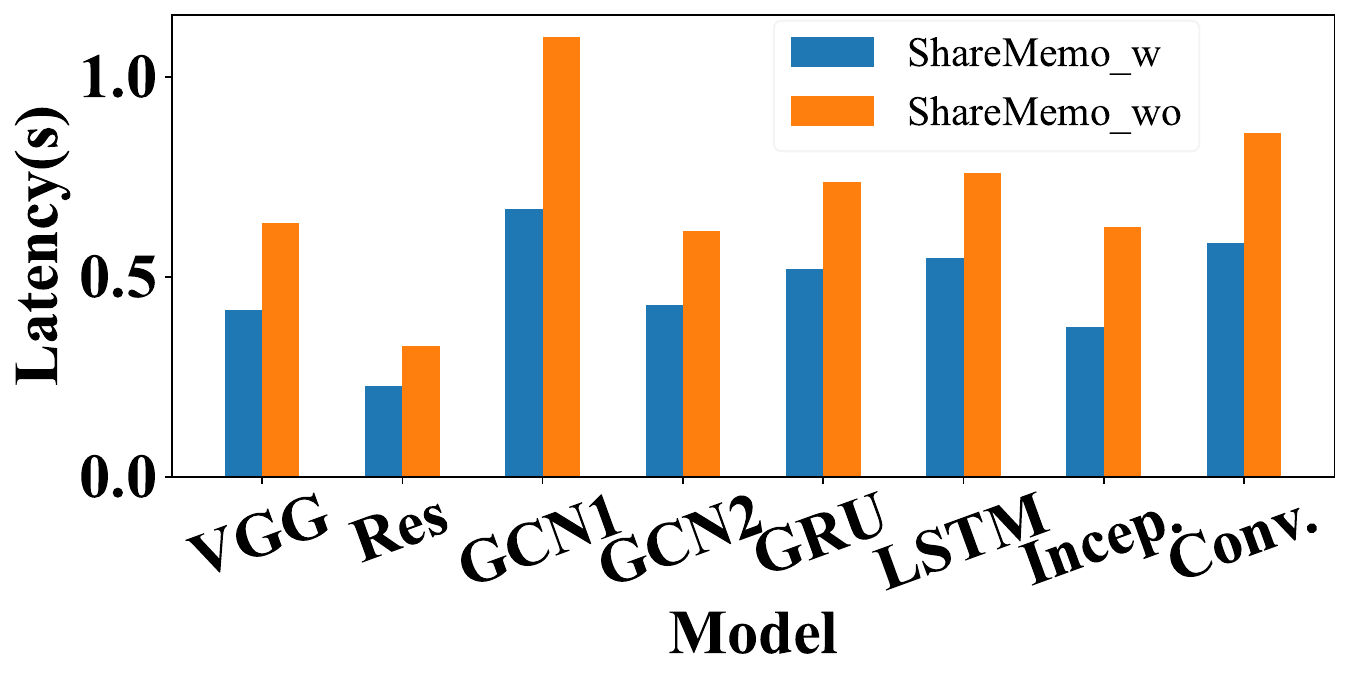}
        \label{fig:3/sharemem}
  }
  \subfloat[AutoEncoder for MOPAR]{
    \includegraphics[scale=0.25]{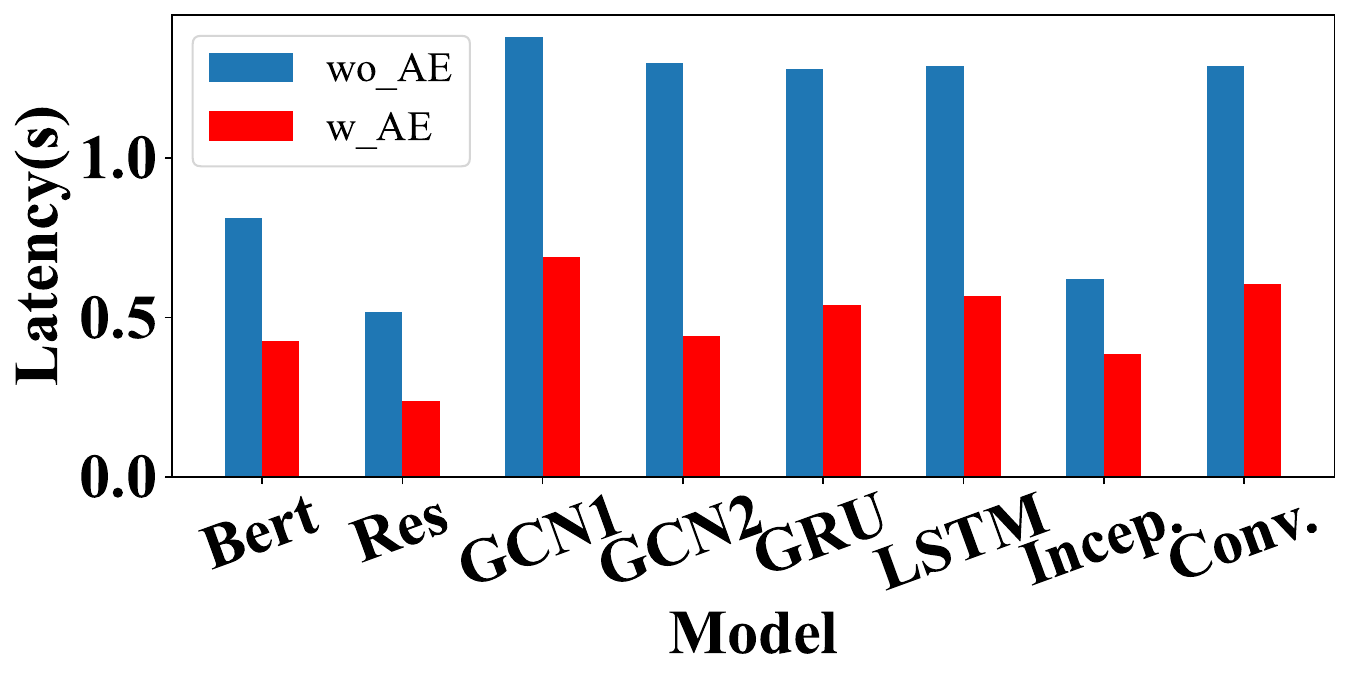}
        \label{fig:3/AE}
  } 
  \vspace{-1mm}
\caption{The effects of MPE and COM on reducing latency.}
\label{fig:iic}
\vspace{-4mm}
\end{figure*}

\subsection{Ablation Studies}
Each component of MOPAR plays a vital role in saving costs and reducing latency. MPE concentrates on improving resource utilization and decreasing execution time, while COM aims to reduce communication overhead. We evaluate their specific contributions by measuring the increased latency or resource usage by ablating different components. Specifically, MPE is replaced with the unsplit strategy, and AE is rendered inoperative by refraining from employing data compression.
We measure resource usage based on memory consumption (MC), which is defined as MC = the amount of allocated memory $\times$ execution time. MC continually assesses the aggregate utilization of resources throughout the execution process \cite{2019small}. 

\noindent
\textbf{MPE:} 
MPE has yielded promising outcomes in terms of memory consumption and inference latency. 
In terms of memory consumption, Figure \ref{fig:memory_footprint} shows that MPE can achieve better performance across the three types of eight DL models, where each rectangle represents the allocated memory quota for each slice. 
Specifically, the average savings on MC for CNNs, RNNs, and GCNs are about 47.46\%, 33.16\%, and 27.97\% respectively. 
Additionally, MPE effectively ensures that the number of slices remains within manageable thresholds. As depicted in Figure \ref{fig:memory_footprint}, each DLIS is partitioned into 3-5 slices, efficiently controlling the overall communication overhead. 

Regarding latency, MOPAR horizontally shards a slice into multiple sub-slices, each of which can run in parallel. Parallelizing slices can shorten the execution time of slices, which in turn reduces the overall latency of DLISs. We validate the effect of MPE on the three types of eight DL models. Figure \ref{fig:3/parallel} shows that enabling MPE in MOPAR significantly reduces latency. 
Overall, MPE reduces the latency of CNNs, RNNs, and GCNs by $1.515 \times$, $1.403 \times $, and $1.273 \times $ on average respectively. For example, disabling MPE increases latency by $1.33 \times $ for VGG and $1.18\times$ for ConvNeXt.

\noindent
\textbf{COM:}
COM has been specifically designed to optimize the cost of slice communication. 
Firstly, we verify the effect of share-memory in COM. As serverless functions lack statefulness, external storage is necessary for data replication and storage. In accordance with the studies \cite{polese2023understanding}, Redis\footnote{https://redis.io/} is employed as the external storage in OpenFaaS. Fig. \ref{fig:3/sharemem} illustrates the latency measurements of the eight DL models, comparing the utilization of Redis with the adoption of share-memory as the communication mechanism. It is observed that the latency of DL models is reduced with share-memory. On average, CNNs, RNNs, and GCNs experience a decrease in latency by 52.01\%, 40.34\%, and 53.15\% respectively. The increase in latency using Redis is primarily attributed to the number of partitioned slices. As DLIS is partitioned into multiple slices, it necessitates more communications with Redis, resulting in higher latency.

Secondly, AE in COM can have a significant impact on reducing communication latency by compressing data.
As depicted in Fig. \ref{fig:3/AE}, data transfer between slices reduces approximately $2.92 \times$ by utilizing compression. Specifically, without AE, the latency of ConvNeXt experiences an increase from 0.605 seconds to 1.29 seconds, signifying a rise of $2.13 \times$. It is worth noting that the other seven models also exhibit an increase in latency when AE is not employed.

\section{Related Work}


\noindent
\textbf{Model Parallelism for Training.}
MOPAR is orthogonal to the extensive body of research on model parallelism in training \cite{li2021terapipe, zhou2022pets}. The process of training and inference entails different stages and objectives. DLISs present a unique range of constraints and opportunities that are not found in training workloads, such as latency. However, the commonality between MOPAR and these methods lies in the fact that their parallel strategies are implemented at different stages.

\noindent
\textbf{Model Inference Systems.} There has been a notable surge in the number of model serving systems in recent years, encompassing a wide spectrum of offerings. These encompass general-purpose production-grade systems such as TorchServe\footnote{https://pytorch.org/serve/}. These systems are employed for the deployment and management of trained models and are widely utilized, although they lack support for restrictions related to SLO or cost.
Furthermore, there exist specialized systems that are optimized for serving a single model \cite{yu2022orca} or cater to specific categories of models, such as transformers \cite{zhou2022pets}. It is worth noting that MOPAR aims to cover a more comprehensive range of models and offers a wider variety of features compared to these existing systems.




AlpaServe \cite{li2023alpaserve} is relevant to our study as it considers both intra- and inter-operator parallelism. 
Nevertheless, the applicability of AlpaServe to serverless environments is hindered by its enumeration strategy, which contradicts the rapid processing capabilities of serverless architectures. Consequently, this misalignment results in higher latency and increased cost.

\noindent
\textbf{Inference on Serverless Computing.} 
Several investigations have examined the provision of inference services in serverless computing \cite{BATCH, INFless}. BARISTA is a scalable system that operates based on the principles of serverless computing and utilizes workload prediction to optimize the allocation of resources for DLISs \cite{BARISTA}. INFless \cite{INFless} achieves high throughput by employing inherent batching and non-uniform scaling mechanisms. However, these approaches overlook the internal characteristics of DLISs, resulting in increased costs. 
To the best of our knowledge, there is a scarcity of studies focusing on DLIS partitioning in serverless computing. 
MOPAR represents a novel endeavor in the realm of serverless computing.

\section{Conclusion}
Under the guidance of ChatGPT, we argue that DLISs will continue to flourish. Serverless computing is a promising solution for DLISs. However, the efficient utilization of resources is of paramount importance for DLISs that serve a large number of requests. In this paper, we initially observe the suboptimal resource efficiency of DLISs due to their resource usage patterns. To address this concern, we design MOPAR, a model partitioning framework that aims to save costs while ensuring latency. Furthermore, we evaluate the efficacy of MOPAR on OpenFaaS and Lambda. One attractive feature of MOPAR is that it is non-intrusive and lightweight, enabling seamless integration into any serverless platform scheduler. In the future, we will focus on optimizing DLISs on GPU-based serverless platforms. In addition, profiling new models based on existing ones presents an interesting direction.

\bibliographystyle{IEEEtranS}
\bibliography{references}

\vfill
\end{document}